\documentclass[aps,prb,twocolumn,floatfix,superscriptaddress]{revtex4-2}
\usepackage[utf8]{inputenc}
\usepackage[T1]{fontenc}
\usepackage[english]{babel}
\usepackage{mathtools}
\usepackage{newtxtext}  % scaled=1.005
\usepackage{amsmath}
\usepackage{amsfonts}
\usepackage[varbb,varg]{newtxmath} % scaled=1.005

\usepackage{booktabs}

\renewcommand\vec{\mathbf}

\usepackage[dvipsnames,svgnames,x11names,hyperref]{xcolor}
\usepackage{hyperref}
\hypersetup{ % closer to what APS journals use
    colorlinks=true, linkcolor=NavyBlue, urlcolor=NavyBlue, citecolor=NavyBlue
}
\usepackage{soul}

\usepackage{physics,braket,siunitx,slashed}
\usepackage[capitalise]{cleveref}

\usepackage{tabularx}

\begin{document}
\title{Microscopic Origin of Orbital Magnetization in Chiral Superconductors}

\author{Jihang Zhu}
\affiliation{Department of Materials Science and Engineering, University of Washington, Seattle, Washington 98195, USA}

\author{Chunli Huang}
\affiliation{Department of Physics and Astronomy, University of Kentucky, Lexington, Kentucky 40506-0055, USA}
\date{\today} 

\begin{abstract}Chiral superconductivity is a time-reversal-symmetry-breaking superconducting phase that has attracted broad interest as a potential platform for topological quantum computation. A fundamental consequence of this symmetry breaking is orbital magnetization, yet a clear microscopic formulation of this quantity has remained elusive. This difficulty arises because Bogoliubov quasiparticles do not carry a definite electric charge, precluding a simple interpretation of orbital magnetization in terms of circulating quasiparticle currents. Moreover, superconductivity and ferromagnetism rarely coexist, and in the few materials where they do (e.g.~uranium-based compounds), strong spin-orbit coupling obscures the orbital contribution to the magnetization. The recent report of chiral superconductivity in rhombohedral multilayer graphene, which has negligible spin-orbit coupling, therefore provides a unique opportunity to develop and test a microscopic theory of orbital magnetization in chiral superconductors. Here we develop such a theory, unifying the interband coherence effects underlying normal-state orbital magnetization with the intrinsic orbital moments of the Cooper-pair condensate. Applying our theory to rhombohedral tetralayer graphene, we find that superconductivity can either enhance or suppress the normal-state orbital magnetization, depending sensitively on the band structure. We further uncover a generalized clapping collective mode that contributes to the magnetization through photon-vertex dressing. Its gap is reduced by the same quantum-geometric form factors that generate the geometric contribution to the superfluid stiffness, and it can be detected optically because the two members of the Cooper pair have unequal energies. Our theory resolves a long-standing conceptual challenge and can be directly tested by nanoSQUID measurements of the orbital magnetization.
\end{abstract}

\maketitle

%\noindent\textsuperscript{*}Correspondence:
%\href{mailto:chunli.huang@uky.edu}{chunli.huang@uky.edu}

%\section*{Summary}Chiral superconductivity is a time-reversal-symmetry-breaking superconducting phase that has attracted broad interest as a potential platform for topological quantum computation.A fundamental consequence of this symmetry breaking is orbital magnetization, yet a clear microscopic formulation of this quantity has remained elusive because Bogoliubov quasiparticles do not carry a definite electric charge. Rhombohedral multilayer graphene, where chiral superconductivity has recently been reported and spin-orbit coupling is negligible, provides a unique opportunity to develop and test such a theory. We formulate orbital magnetization by unifying the interband coherence underlying normal-state magnetization with the intrinsic orbital moments of the Cooper-pair condensate. Applied to rhombohedral tetralayer graphene, our theory shows that superconductivity can either enhance or suppress normal-state orbital magnetization, depending  on the band structure. It also predicts optically active generalized clapping modes whose gaps are reduced by quantum-geometric effects. These results resolve a long-standing conceptual challenge and provide predictions directly testable by nanoscale superconducting quantum interference device measurements.

%\textbf{Keywords:} chiral superconductivity, topological superconductivity, orbital magnetization, rhombohedral graphene, Cooper-pair orbital moment, quantum geometry, clapping mode, nanoSQUID magnetometry.

\section{Introduction}
The Bardeen-Cooper-Schrieffer (BCS) theory \cite{bardeen1957theory}, originally formulated for time-reversal-invariant superconductors, was soon generalized by Anderson and Morel \cite{Anderson_Morel_1961} to higher-angular-momentum pairing states, including chiral phases that spontaneously break time-reversal symmetry. Despite decades of theoretical effort~\cite{mcclure1979angular,stone2004edge,braude2006orbital,sauls2011surface,tada2015orbital,huang2014vanishing,etter2018spontaneous,robbins2020theory,xiao2021conserved}, however, formulating orbital magnetization in chiral superconductors at a microscopic level has proven conceptually challenging.
One major reason is that much of the early theoretical understanding of orbital angular momentum in chiral superfluids, such as the 
$^3$He-A phase, was developed in the context of translationally and Galilean-invariant systems. In such systems, spatial inhomogeneity arises only at sample boundaries where the order parameter necessarily vanishes, and boundary physics plays a subtle role in determining the orbital angular momentum and the associated edge currents~\cite{sauls2011surface,stone2004edge,tada2015orbital,huang2014vanishing}.
In crystalline superconductors, by contrast, Galilean invariance is explicitly broken by the periodic lattice potential, giving rise to an intrinsic bulk contribution to the orbital response. In particular, the velocity operator of crystalline material is not simply $\mathbf{p}/m$ but consists of both the intraband Fermi velocity, $\nabla_{\mathbf{k}} \epsilon_F({\mathbf{k}})/\hbar$, and the interband velocity matrix elements
between Bloch states 
$\braket{u_{c\mathbf{k}}|\mathbf{v}|u_{v\mathbf{k}}}$. The interband matrix-elements are known to be essential in generating orbital magnetization even in normal (nonsuperconducting) crystals~\cite{Ceresoli_OM_2006,souza2008dichroic,xiao2007valley,shi2007quantum,xiao2010berry,zhu2025magnetic}, making it natural and necessary to account for them in the superconducting phase.

\begin{figure}[t]
    \centering
\includegraphics[width=0.9\linewidth]{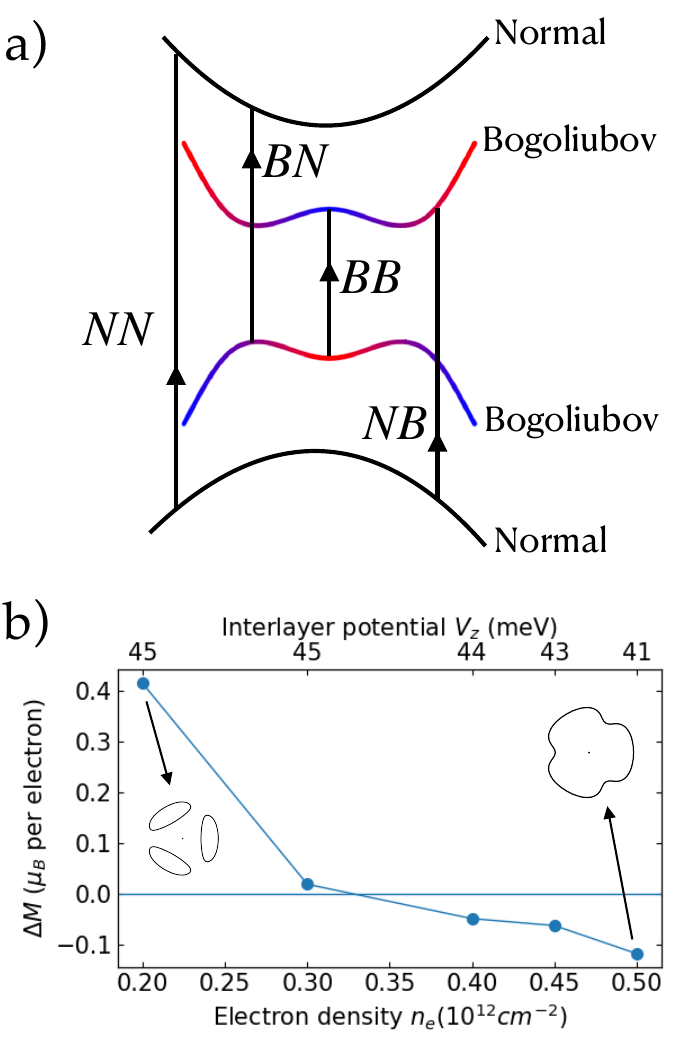}
\caption{(a) Schematic of interband processes contributing to orbital magnetization in a chiral superconductor. The middle two (Bogoliubov) bands are superconducting, while the others remain normal. Transitions are classified by whether the initial and final states are normal or Bogoliubov (see Sec.~III). (b) Superconductivity-induced change in orbital magnetization, $\Delta M = M_z^{\rm S} - M_z^{\rm QM}$, across a Lifshitz transition as carrier density and interlayer potential are tuned. Inset: Fermi-surface topology of the quarter-metal phase.} %The calculation is done for rhombohedral tetralayer graphene in valley $K$.
\label{fig:schematic}
\end{figure}

This raises a fundamental question: how should we formulate orbital magnetization in a superconductor to consistently incorporate both the intrinsic orbital moment of Cooper pairs, which arises from pairing within a narrow energy window around the Fermi surface, and the interband coherence induced by the periodic crystal potential, whose characteristic energy scales are typically much larger than the superconducting gap? Here we develop a microscopic theory that treats these two contributions on equal footing by moving beyond the traditional Fermi-surface-only description and systematically accounting for all virtual transitions between normal (Landau) quasiparticles and Bogoliubov quasiparticles (Fig.~\ref{fig:schematic}a).

%The logic of our formulation is as follows. We compute the orbital magnetization from the linear response of the superconducting state to a slowly varying external vector potential. This requires identifying the dressed photon vertex, namely the physical electromagnetic current with vertex correction. A key point is that this vertex is not identical to the bare Bogoliubov quasiparticle velocity obtained by differentiating the Bogoliubov-de Gennes Hamiltonian with respect to momentum. The quasiparticle velocity describes the propagation of Bogolons, whereas the dressed photon vertex determines how the superconducting state couples to the external vector potential. The dressing of this vertex incorporates collective superconducting degrees of freedom and is required for gauge invariance and current conservation. The distinction between quasiparticle velocity and dressed photon vertex is essential for obtaining the microscopic orbital magnetization in a superconductor.

We apply this theory to chiral superconductivity in rhombohedral tetralayer graphene, focusing on the quarter-metal phase~\cite{han2408signatures}, whose band structure is well characterized. This phase exhibits a strong anomalous Hall effect~\cite{zhou2021half}, and its orbital magnetization has been directly visualized via nanoSQUID measurements~\cite{arp2024intervalley,auerbach2025isospin}. 
The central question, therefore, is how this orbital magnetization changes upon entering the superconducting state. We find that the effect of superconductivity on orbital magnetization depends sensitively on the normal-state bandstructure. Using a phenomenological attractive p-wave pairing model for rhombohedral tetralayer graphene, we show that superconductivity enhances the orbital magnetization when the quarter metal hosts three disjoint Fermi pockets, but suppresses it when the Fermi surface is simply connected (Fig.~\ref{fig:schematic}b). This contrast yields experimentally testable predictions accessible through combined nanoSQUID and quantum oscillation measurements.

\subsection{Lessons from normal-state orbital magnetization}
    Before turning to the superconducting state, it is useful to step back and discuss how orbital magnetization is formulated in normal state. We begin by emphasizing a basic but fundamental point: spontaneous orbital magnetization, like spin ferromagnetism, is an interaction-driven phenomenon because the system must choose a time-reversal-breaking state from a microscopic time-reversal invariant Hamiltonian. While the role of interactions in spin ferromagnetism is relatively transparent \cite{herring1966exchange}, for example through exchange interactions that favor spin polarization, their role in orbital magnetism is more subtle. Interactions must somehow select one pattern of circulating currents, over its time-reversed partner. In dilute K-valley 2D materials such as graphene and transition-metal dichalcogenides, the valley degree of freedom provides a transparent route to spontaneous orbital magnetization. Because opposite valleys can carry opposite orbital magnetic moments, valley polarization can produce a net orbital magnetization. 
    
    More generally, however, the modern theory of orbital magnetization is often formulated using an effective single-particle Hamiltonian $\tilde H$ that already breaks time-reversal symmetry, without explicitly keeping track of the interactions that generated this symmetry breaking. One then defines the current operator from $\tilde H$ and computes the orbital magnetization from the corresponding single-particle eigenstates. This raises a conceptual question: what is the appropriate current operator, given that the $\tilde H$  contains interaction-generated mean-field contributions?

This issue was recently clarified in Refs.~\cite{XLiu_OMHF_2025, kang2025orbital, zhu2025magnetic}. In our work \cite{zhu2025magnetic}, the starting point is a many-body Hamiltonian that is time-reversal symmetric. The system is coupled to an external vector potential in the velocity gauge,
$H'=-e\vec{v}\cdot \vec A$ where $\vec v$ is the physical velocity operator of the microscopic Hamiltonian and contains no explicit interaction contribution. When this perturbation acts on the mean-field ground state, it mixes occupied orbitals $|u_{v \vec k}\rangle$ with unoccupied orbitals $|u_{c\vec k}\rangle$, with coefficients of the form
$\frac{\vec{\Gamma}_{v\vec k,c\vec k}\cdot \vec A_{\vec q}}
{\epsilon_{v\vec k}-\epsilon_{c\vec k}}$.
This structure is expected from perturbation theory, but the crucial point is that the vertex $\vec{\Gamma}_{v\vec k,c\vec k}$ is the velocity vertex of the mean-field Hamiltonian, as required by the Ward identity. Thus, although the microscopic coupling to the electromagnetic field involves the bare physical velocity $v$, the response of the interacting broken-symmetry state is governed by the properly dressed mean-field velocity vertex $\Gamma$.
This lesson is important for superconductors. Consider a BCS mean-field Hamiltonian,
\begin{equation}
H(\vec k)
=
\begin{pmatrix}
h(\vec k)-\mu & \Delta(\vec k) \\
\Delta^\dagger(\vec k) & -h^T(-\mathbf k)+\mu
\end{pmatrix}.
\label{eq:BdG_H}
\end{equation}
What is the correct current operator to use in defining loop currents and orbital magnetization? A natural first guess is
\begin{equation}
    v_1= \frac{1}{\hbar}\nabla_k\begin{pmatrix}
h(\mathbf k) & 0 \\
0 & h^T(-\mathbf k)
\end{pmatrix}.
\end{equation}
However, this choice is unsatisfactory because it does not include the mean-field self-energy. From the normal-state lesson above, one expects the electromagnetic response to involve the velocity vertex of the appropriate mean-field Hamiltonian. If we couple to 
\begin{equation}
    v_2= \frac{1}{\hbar}\nabla_k\begin{pmatrix}
h(\mathbf k)-\mu & \Delta(\mathbf k) \\
\Delta^\dagger(\mathbf k) & -h^T(-\mathbf k)+\mu
\end{pmatrix},
\end{equation}
this choice is also problematic, because Bogoliubov quasiparticle flow is not the same as conserved electric charge flow~\cite{parameswaran2012microscopic}. The strategy of this work is therefore to formulate superconducting orbital magnetization directly from the electromagnetic response of the underlying many-body Hamiltonian, rather than guessing a current operator at the mean-field level. Indeed, our result shows that the magnetization cannot be determined from the single-bologon spectrum (BCS mean-field spectrum) alone. Two-Bologon processes that give rise vertex corrections (and collective modes) are also required. This is the key difference from the normal-state orbital magnetization.

\section{Methods}
Previous theoretical studies~\cite{xiao2021conserved,annett2009orbital,Robbins2019Anomalous} have provided insights into orbital magnetization in superconductors, often through semiclassical quasiparticle dynamics or expectation values of angular-momentum operators such as $\mathbf r\times \mathbf p$. While these approaches are physically transparent in the normal state, their extension to superconductors is subtle. The relation between angular momentum and orbital magnetization is no longer fixed by a simple gyromagnetic factor~\cite{annett2009orbital,Robbins2019Anomalous}.
It is also unclear whether semiclassical formulations ~\cite{xiao2021conserved} fully account for vertex corrections, because a Bogoliubov quasiparticle wave packet is incompatible with a strictly uniform order parameter: it necessarily induces a spatial deformation of the condensate, generating a dipolar backflow that extends over distances of order the London penetration depth~\cite{parameswaran2012microscopic}.

To avoid the above complications, we instead begin from the thermodynamic definition of orbital magnetization $M=-d\Omega/dB*1/\text{Area}$. Consider a generic microscopic $k\!\cdot\!p$ continuum Hamiltonian, expanded to linear order in vector potential:
\begin{align}\label{eq:H_micro}
    H &=
    \sum_{\vec{k}\vec{q} \alpha\beta}
    \left[
    h_{\alpha\beta}(\vec k)\delta_{\vec q,0}
    +
    \frac{e}{2}
    \big(
        \vec v_{\alpha\beta}(\vec k)
        + \vec v_{\alpha\beta}(\vec k+\vec q)
    \big)\!\cdot\!\vec A_{\vec q}\right]
    c_{\vec{k}+\vec q,\alpha}^\dagger c_{\vec{k}\beta}
    \nonumber\\
    &\quad
    + \frac{1}{2}
    \sum_{\vec{k}\vec{k}'\vec q\alpha\beta}
    V_{\vec{k},\vec{k}'}
    c_{\vec{k}+\frac{\vec q}{2},\alpha}^\dagger
    c_{-\vec{k}+\frac{\vec q}{2},\beta}^\dagger
    c_{-\vec{k}'+\frac{\vec q}{2},\beta}
    c_{\vec{k}'+\frac{\vec q}{2},\alpha}.
\end{align}
Here $h_{\alpha\beta}(\vec{k})$ are matrix elements of a continuum $k\!\cdot\!p$ Hamiltonian and  
$
    \vec v(\vec{k}) = \nabla_{\vec p} h(\vec p)\big|_{\vec p=\vec k}/\hbar
$
represents the group-velocity operator of the Bloch waves \footnote{Here we keep only the terms linear in $\mathbf A$, which are sufficient for extracting the spontaneous orbital magnetization from the term in $\Omega$ linear in $B$. Field-induced magnetization and orbital susceptibility require the expansion of the microscopic Hamiltonian to quadratic order in $\mathbf A$, including the usual diamagnetic terms.}. We apply minimal coupling to the Hamiltonian by introducing a vector potential
\begin{equation}
    \vec A(\hat{\vec r})=\sum_{\vec q}\vec A_{\vec q}\,e^{i\vec q\cdot \hat{\vec r}}
\end{equation}
Here $\hat{\vec r}$ is the position operator, $\alpha,\beta$ label orbital degrees of freedom of the electrons. For clarity, we consider spinless fermions and pairing at zero center-of-mass momentum.

%Extensions to spinful fermions and finite center-of-mass pairing are straightforward but would unnecessarily complicate the notation.
We then define the Nambu spinor
\begin{equation}
       \Psi_{\vec k} =
    \begin{bmatrix}
        c_{\vec k\alpha_1} & \dots & c_{\vec k\alpha_N} &
        c_{-\vec k\alpha_1}^\dagger & \dots & c_{-\vec k\alpha_N}^\dagger
    \end{bmatrix}^T.
\end{equation}
In this basis, the mean-field Hamiltonian can be written as
\begin{equation}
    H = \frac{1}{2} \sum_{\vec k,\vec q}
    \Psi_{\vec k+ \vec q}^\dagger
    \left(
    H^{\rm BCS}({\vec k})\,\delta_{\vec q,0} 
    +e \,\vec A_{\vec q} \cdot \vec{\Gamma}_{\vec k}^0(-\vec q)
    \right)  
    \Psi_{\vec k},
\end{equation}
where the BCS mean-field Hamiltonian is
\begin{equation}
    H^{\rm BCS}(\vec k) =
\begin{pmatrix}
h(\vec k) & \Delta(\vec k)  \\
\Delta^\dagger(\vec k) & -h^T(-\vec k) 
\end{pmatrix},
\end{equation}
and the pairing potential is
$
    \Delta_{\alpha\beta}(\vec{k}) =
    \sum_{\vec{k}'} V_{\vec{k},\vec{k}'}
    \langle c_{-\vec{k}'\beta} c_{\vec{k}'\alpha} \rangle
$. The photon-vertex is given by
\begin{equation}\label{eq:Gamma_0}
    \vec{\Gamma}_{\vec k}^0(-\vec q)=
\begin{pmatrix}
\tfrac{1}{2}\big(\vec v(\vec k)+ \vec v(\vec k+\vec q)\big) & 0  \\
0 & -\tfrac{1}{2}\big(\vec v^T(-\vec k)+ \vec v^T(-\vec k-\vec q)\big)
\end{pmatrix}.
\end{equation}
Both the normal-state density matrix and the anomalous pairing amplitude can be described together using the quasiparticle density matrix
%To make the connection with conventional Hartree-Fock theory explicit, we describe both the normal-state density matrix $\langle c_{\vec{k}\beta}^\dagger c_{\vec{k}\alpha}\rangle$ and the anomalous pairing amplitude $\langle c_{-\vec{k}\beta} c_{\vec{k}\alpha}\rangle$ in a unified manner using the quasiparticle density matrix 
$P = \sum_{\vec k} P({\vec k})$, $ P_{ij}(\vec k) =
    \langle \Psi_{\vec k j}^\dagger \Psi_{\vec k i} \rangle $
where $i,j$ label components of the Nambu spinor.  At self-consistency, the quasiparticle density matrix $P$ satisfies the stationarity condition
$[H^{\mathrm{BCS}}({\vec k}), P({\vec k})] = 0 $, $\forall\,\vec k,$
and takes the form of a projector onto the occupied states,
\begin{equation}
     P({\vec k}) =\sum_{i}
\ket{U_{i\vec{k}}}\bra{U_{i\vec{k}}}\;,\; H^{\mathrm{BCS}}({\vec k})\ket{U_{i\vec{k}}}=E_{i\vec{k}}\ket{U_{i\vec{k}}}.
\end{equation}
Throughout this work we use the notation, $ i,j$ for occupied bands and $m,n$ for unoccupied bands. We caution the reader that the term ``occupied state'' in this context does not imply a fixed particle number, i.e.~$\sum_{\vec k}\mathrm{Tr}(P({\vec k}))$ is not the total number of electrons because the BCS ground state is not a charge eigenstate. This language is used only to emphasize the analogy with the Hartree-Fock density matrix.

We now study the change of the ground state to linear order in the vector potential $\vec{A}_{\vec{q}}$ and use the perturbed ground state to evaluate the leading field-dependent correction to the energy, where $\vec{q}\times\vec{A}_\vec q=i\vec{B}_{\vec{q}}$. Note we used a periodic magnetic field with a finite wave vector $\vec{q}$ and take the limit $\vec{q}\rightarrow0$ only at the end of the calculation. Introducing a uniform magnetic field from the beginning would require a vector potential that grows without bound  (e.g.~$\vec A=\vec{B}\times\vec{r}/2$) and change the continuous spectrum to closely spaced Landau levels. The corresponding wave functions would varies over magnetic length $\ell_B\propto B^{-1/2}$, which diverges as $B\rightarrow0$ and eventually becomes comparable to the system size. 
 The electron orbits then become sensitive to the sample boundary, making the weak-field limit particularly subtle. This boundary sensitivity is the main difficulty in the early theory of orbital magnetism, as discussed nicely in Sec.~4.3 of Ref.~\cite{peierls2020surprises}.
 The finite-$q$ perturbative calculation avoids these difficulties by evaluating the response at nonzero $\vec{q}$ before taking the long-wavelength limit. This method goes back to the early days of quantum many body theory to compute orbital susceptibility of a time-reversal invariant state \cite{hebborn1964orbital} and was subsequently used to calculate spontaneous orbital magnetism of a time-reversal broken state in Refs.~\cite{shi2007quantum,vignale2026ward}. 
 In Ref.~\cite{huang2026orbital}, we use a simple toy model to  show that the finite-$q$ perturbative approach and total energy calculation calculated with unbounded vector potential (e.g. $\vec A=\vec{B}\times\vec{r}/2$) yield the same orbital magnetization. Motivated by this agreement in the normal state, we now apply the finite-$q$ approach to a superconductor.

We write the perturbed density matrix as follows,
\begin{equation}
\delta P =
\sum_{\vec k,m,i} X_{m\vec k+\vec q, i\vec k} \, |U_{m \vec k+\vec q}\rangle \langle U_{i \vec k}|
+ X_{i \vec k- \vec q, m \vec k} \, |U_{i \vec k-\vec q}\rangle \langle U_{m \vec k}|.
\end{equation}
where $X\propto\vec A_{\vec q}$ are linear-response coefficients. The resulting change in the grand potential is $ \delta \Omega = \mathrm{Tr}(\delta P\,\hat{\Omega})$,
\begin{align} 
\delta \Omega =\frac{1}{2} \sum_{\vec kmi}  X_{m \vec k+\vec q, i \vec k} \langle U_{i \vec k} | \hat{\Omega} | U_{m \vec k+\vec q} \rangle + X_{i \vec k-\vec q, m \vec k} \langle U_{m \vec k} | \hat{\Omega} | U_{i \vec k-\vec q} \rangle. 
\end{align}
Since the magnetic field is linear order in both $\vec A$ and $\vec q$, we expand $\delta\Omega$ to first order in these quantities. Because the matrix elements
$\langle U_{i \vec k} | \hat{\Omega} | U_{m \vec k+\vec q} \rangle$
vanish at $\vec q=0$, the leading $O(Aq)$ contribution to $\delta\Omega$ must arise from the linear-in-$q$ term of the expansion,
 $
    \langle U_{i \vec k} | \hat{\Omega} | U_{m \vec k+\vec q} \rangle = 
    q_\mu
    \langle U_{i\vec k}| \partial_\mu U_{m \vec k} \rangle 
    (E_{i \vec k} + E_{m \vec k})/2 +O(q^2)
$. Retaining only this leading order, $\delta \Omega$ becomes
\begin{align} \label{eq:delta_Omega} 
\delta \Omega &=\frac{1}{2}\sum_{\vec k m i} (E_{m \vec k} + E_{i \vec k}) q_\mu \mathrm{Re}\!\left[ X_{m \vec k, i \vec k} \langle U_{i \vec k} | \partial_\mu U_{m \vec k} \rangle \right] + O(q^2)
\end{align}
and therefore only the coefficients $X$ evaluated at $\vec q=0$ are required.
These coefficients are determined by the following linear equation from many-body perturbation theory \cite{thouless2013quantum,zhu2025magnetic}:
\begin{equation} \label{eq:X_linearized}
   \big(E_{m\vec k}-E_{i\vec{k}}\big)\, X_{m\vec k , i\vec k}
   -\delta \Sigma_{m\vec k, i \vec k}
   =-e \,\bra{U_{m\vec k}}\vec{\Gamma}_{\vec{k}}^0\ket{U_{i\vec{k}}}\cdot \vec A,
\end{equation}
where $\vec{\Gamma}_{\vec{k}}^0$ is the bare photon vertex, obtained by setting $\vec q=0$ in Eq.~\eqref{eq:Gamma_0}, and $\delta\Sigma$ is the induced self-energy. This equation is a coupled linear system, since the perturbed pairing self-energy $\delta \Sigma$ generates transitions from other occupied states $\ket{U_{j\vec k'}}$ to unoccupied states $\ket{U_{n\vec k'}}$ with amplitudes $X_{n\vec k',j\vec k'}$. Solving this system is equivalent to carrying out the vertex correction, i.e., a summation of self-energy insertions into the photon vertex. The solution can be written in the following compact form,
\begin{equation}\label{eq:X_matrix}
     X_{m\vec k, i\vec k}
     =\frac{-e \bra{U_{m\vec k}}\vec{\Gamma}_{\vec{k}}\ket{U_{i\vec{k}}}\cdot \vec A}
     {E_{m\vec k}-E_{i\vec{k}}},
\end{equation}
where $\vec{\Gamma}_{\vec k}$ is the dressed photon vertex. Substituting this expression into Eq.~\eqref{eq:delta_Omega}, choosing a gauge with $
    A_\nu q_\mu 
    = i\epsilon_{\nu\mu\lambda} B_\lambda/2,$
and using the thermodynamic definition of the orbital magnetization $
    M_\lambda = -(\text{Area})^{-1}
    \partial\Omega/\partial B_\lambda$,
we obtain
\begin{equation} 
\begin{split}
M_\lambda =&\frac{e}{4} \int \frac{d^2 \vec k}{4\pi^2}\sum_{m i} \frac{E_{m \vec k} + E_{i \vec k}}{E_{m \vec k} - E_{i \vec k}} \epsilon_{\lambda \mu \nu}\\& \mathrm{Im} \left[ \bra{U_{m\vec k}} \Gamma^{\nu}_{\vec k}\ket{U_{i\vec k}} \langle U_{i \vec k} | \partial_\mu U_{m \vec k} \rangle \right]. 
\end{split}
\end{equation}

\begin{table}[t]
\centering
\renewcommand{\arraystretch}{1.25}
\begin{tabular}{lcc}
\textbf{Operator} & \textbf{Normal state} & \textbf{Superconducting state} \\
\hline

Bare velocity 
& $\vec v(\vec k)=\frac{1}{\hbar}\nabla_{\vec k} h(\vec k)$ 
& -- \\[4pt]

Quasiparticle velocity 
& $\frac{1}{\hbar}\nabla_{\vec k} H^{\rm HF}(\vec k)$ 
& $\frac{1}{\hbar}\nabla_{\vec k}
\begin{pmatrix}
h(\vec k) & \Delta_{\vec k} \\
\Delta^\dagger_{\vec k} & -h^{T}(-\vec k)
\end{pmatrix}$ \\[6pt]

Bare photon vertex 
& $\vec v(\vec k)$ 
& $\vec{\Gamma}_{\vec k}^{\,0}=\frac{1}{\hbar}
\begin{pmatrix}
\nabla_{\vec k} h(\vec k) & 0 \\
0 & \nabla_{\vec k} h^{T}(-\vec k)
\end{pmatrix}$ \\[6pt]

Dressed photon vertex 
& $\frac{1}{\hbar}\nabla_{\vec k} H^{\rm HF}(\vec k)$ 
& $\vec{\Gamma}_{\vec k}=
\bigl(1+\tfrac{C}{\Delta E}\bigr)^{-1}\vec{\Gamma}_{\vec k}^{\,0}$ \\

\hline
\end{tabular}
\caption{\label{tab:velocities} Comparison of velocity and photon–vertex operators in the normal and superconducting states.
In the normal state, the quasiparticle velocity and the dressed photon vertex coincide, reflecting charge conservation for Landau quasiparticles. In the superconducting state, these operators become distinct because Bogoliubov quasiparticles are not charge eigenstates. Here $H^{\rm HF}(\vec k)=h(\vec k)+\Sigma^{\rm HF}(\vec k)$  is Hartree–Fock Hamiltonian and $\vec{\Gamma}_{\vec k}$ is the dressed vertex [Eq.\eqref{eq:vertex}]. }
\end{table}
Since $\langle U_{i \vec k}| H^{\rm BCS}_\vec k| U_{m \vec k} \rangle=E_{m\vec k} \langle U_{i \vec k} |U_{m \vec k}\rangle=0$,
we can always relate the interband Berry connection to matrix elements of the mean-field quasiparticle velocity:
\begin{equation}
    \langle U_{i \vec k} | \partial_\mu U_{m \vec k} \rangle
    =-\frac{
    \langle U_{i \vec k} | (\partial_\mu H^{\rm BCS}({\vec k}))| U_{m\vec k} \rangle
    }{E_{i\vec k}-E_{m\vec k}}.
\end{equation}
\begin{table}[t]
\centering
\renewcommand{\arraystretch}{1.25}
\begin{tabular}{lc}
\hline
$(m,i)$ & \textbf{Orbital magnetization in a superconductor} \\
\hline
$(N,N)$ 
  & Inherited from the parent normal state
  ($M^{\rm NN}$) \\[2pt]
\hline
\begin{tabular}[c]{@{}l@{}}
$(N,B)$ \\
$(B,N)$
\end{tabular}
  & Normal–Bogoliubov mixing  ($M^{\rm NB}$, $M^{\rm BN}$) \\[4pt]
\hline
$(B,B)$ 
  & Cooper-pair contribution ($M^{\rm BB}$) \\[6pt]
\hline 
\end{tabular}
\caption{\label{tab:M_classification}
Classification of orbital magnetization in a superconductor. $(m,i)$ label unoccupied and occupied states in Eq.~\eqref{eq:M_general}. $N$ and $B$ denote normal (Landau) quasiparticles and Bogoliubov quasiparticles.
}
\end{table}
This leads to our main result, which expresses $M_\lambda$ in terms of the BCS quasiparticle energies and wave functions:
\begin{align}  \label{eq:M_general}
M_\lambda =& \frac{e}{4} \int \frac{d^2k}{(2\pi)^2} \sum_{ m i} \frac{E_{m \vec k} + E_{i \vec k}}{(E_{m \vec k} - E_{i \vec k})^2} \epsilon_{\lambda \mu \nu}\times \nonumber \\ &\mathrm{Im} \left( \bra{U_{m\vec k}} \Gamma^{\nu}_{\vec k}\ket{U_{i\vec k}} \langle U_{i \vec k} | (\partial_\mu H^{\rm BCS}(\vec k))| U_{m \vec k} \rangle \right). 
\end{align}

This form of the orbital magnetization is very illuminating. The quantity $M_\lambda$ depends on two types of transition matrix elements between occupied states $i$ and unoccupied states $m$. The first is the dressed photon-vertex $\Gamma^\nu_{\vec k}$ which determines the electromagnetic response like the Meissner effect. The second is the Bogoliubov quasiparticle group-velocity operator $\partial_\mu H^{\rm BCS}_{\vec k}$, which describes the propagation of quasiparticles in the superconducting state. 

The important point is that, unlike in the normal state, Bogoliubov quasiparticles are not charge eigenstates. As a result, the group-velocity operator does not coincide with the photon vertex. By contrast, in the normal (non-superconducting) state, the solution of the linearized mean-field equation yields a photon vertex of the form $\Gamma^\nu_{\vec k} = \hbar^{-1}\partial_\nu H^{\rm HF}_{\vec k}$~\cite{zhu2025magnetic}, which is identical to the quasiparticle group velocity. Substituting this relation and replacing $H^{\rm BCS}$ with $H^{\rm HF}$, our expression reduces to the standard normal-state formula for orbital magnetization obtained from semiclassical theory. Physically, when Landau quasiparticles flow with a given velocity, they necessarily carry a corresponding charge current because they are charge eigenstates. This relationship is broken in the superconductor when Landau quasiparticles become Bogoliubov quasiparticles. A summary of these distinctions is in Table.~\ref{tab:velocities}.   Since superconductivity typically develops only in a subset of bands in a periodic crystal, it is convenient to classify the orbital magnetization $M$ according to Table~\ref{tab:M_classification} and write the total magnetization as,
\begin{equation}
    M_\lambda =
    \int \frac{d^2 \vec k}{4\pi^2}
    %\sum_{\vec k}
    \Big[
    M^{\rm NN}_\lambda(\vec k) + M^{\rm NB}_\lambda(\vec k) + M^{\rm BN}_\lambda(\vec k) + M^{\rm BB}_\lambda(\vec k)
    \Big].
\end{equation}
%These contributions are illustrated in Fig.~\ref{fig:schematic}a.

%For a superconductor with time-reversal symmetry, the quasiparticle energy $E_{n\vec k}=E_{n-\vec k}$, and both the photon vertex and the velocity operator are odd under time reversal. However, because time-reversal symmetry is antiunitary, it also complex conjugates the matrix elements. Writing the quantity inside the imaginary part of Eq.~\eqref{eq:M_general}, as $f(\vec k)$, time reversal symmetry implies $\Im(f(\vec{k}))= \Im(f^*(-\vec k))$. Consequently, the integrand is an odd function of $\vec k$ and vanishes upon integration over the whole Brillouin zone. %Thus, the orbital magnetization vanishes in time-reversal symmetric superconducting state.
 
%I suspect some formula in the literature can be recovered by substituting $\Gamma^\nu = \partial_\nu H^{\rm BCS}+(\Gamma^\nu-\partial_\nu H^{\rm BCS})$ into the formula above and separate M into two contributions $M=M^{(1)}+M^{corr}$.-------

\begin{figure*}[t]
    \centering
\includegraphics[width=0.95\linewidth]{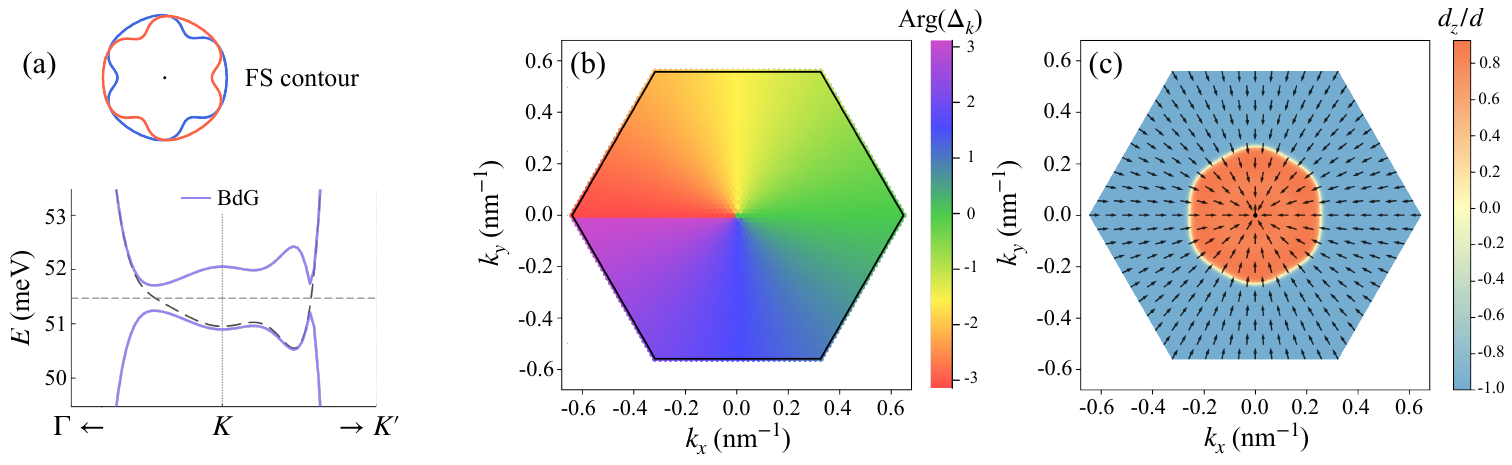}
\caption{
Superconducting ground state of rhombohedral tetralayer graphene in valley $K$ at $V_z=41$~meV and carrier density $n_e=0.5\times10^{12}$~cm$^{-2}$, corresponding to the highest-density point shown in Fig.~\ref{fig:schematic}b). 
(a) Fermi-surface contours of $\xi_{c,\mathbf{k}}$ (blue) and $\xi_{c,-\mathbf{k}}$ (orange), together with the self-consistent Bogoliubov--de Gennes spectrum. 
The black dashed curve shows the normal-state conduction-band dispersion, and the gray dashed horizontal line indicates the Fermi level.
(b) Phase of the superconducting order parameter $\Delta_{\mathbf{k}}$.
(c) Skyrmion-like texture of the pseudospin field $\vec d_{\vec k}$, with the in-plane components $(d_x,d_y)$ shown as arrows and the out-of-plane component $d_z$ encoded in the color scale. 
The in-plane components are normalized by $(d_x^2+d_y^2)^{1/2}$, and $d_z$ is normalized by the total magnitude $d=(d_x^2+d_y^2+d_z^2)^{1/2}$. 
%All calculations are done at $g=160$~meV$\cdot$nm$^2$, chosen to reproduce the experimentally observed critical temperature.
}
%(b) and (c) reveal the chiral structure of the superconducting order. Valley $K$ and $K'$ spontaneously select superconducting order parameter of opposite chirality, as evidenced by the textures in \jz{}{supp...}. The total energy per particle in the superconducting ground state is 50.4944 meV, and the energy splitting between the two chiral states is small, only 0.0444 meV per electron.
    % The calculation is performed using a cutoff momentum $k_{cf} = 0.038b \approx 0.647$ nm$^{-1}$, where $b = 4\pi a_0/\sqrt{3}$ is the reciprocal lattice vector.
\label{fig:mf_solution}
\end{figure*}

\subsection{Chiral superconductors in rhombohedral multilayer graphene}

Since the initial observation of superconductivity in rhombohedral trilayer graphene~\cite{zhou2021superconductivity}, many more superconducting phases has been identified in rhombohedral multilayer graphene~\cite{zhou2022isospin,zhang2023enhanced,patterson2025superconductivity,han2025signatures,seo2025family,morissette2025superconductivity,kumar2025pervasive,guo2025flat,xie2025magnetic}. Several of these phases are induced or enhanced by external magnetic fields~\cite{zhou2022isospin,seo2025family,kumar2025pervasive,guo2025flat,xie2025magnetic}, suggesting spin-polarized order-parameter, while others display unconventional directional responses~\cite{morissette2025superconductivity}. 
Although the theoretical framework developed here applies broadly to these systems, we focus mainly here on the chiral superconductor  whose parent state is an unambiguously symmetry-broken quarter metal with anomalous Hall effect. This is because this realizes a simple theoretical model of a single-component fermion, involving only a single valley, a single spin projection, and a single conduction band. A unique advantage of graphene-based superconductors compared to other candidate chiral superconductors is their relatively well understood electronic band structure, e.g, see Extended data Table 1 of Ref.~\cite{zhou2023imaging}. For rhombohedral $N$-layer graphene, the low-energy electronic structure near a given valley is described by a $k\!\cdot\!p$ Hamiltonian with $2N$ bands,
\begin{equation}
h^{k\cdot p}=\sum_{n,\vec k}\xi_{n\vec k}\ket{\phi_{n\vec k}}\bra{\phi_{n\vec k}}, 
\qquad
\xi_{n\vec k}=\epsilon_{n\vec k}-\mu .
\end{equation}
We assume that superconductivity develops in the first conduction band label as $c$, and that the attractive interaction is much smaller than the spectral separation to other bands, see Appendix. Projecting onto this band yields the effective Hamiltonian
\begin{equation}
H_c = \sum_{k}\xi_{c \vec k}
a^\dagger_{c\vec k} a_{c\vec k}+
\frac{1}{2} \sum\limits_{\vec k,\vec k'} \tilde{V}_{\vec k,\vec k'} a^\dagger_{c\vec k} a^\dagger_{c,-\vec k} a_{c,-\vec k'} a_{c\vec k'}
\end{equation}
\begin{equation}
    \tilde{V}_{\vec k,\vec k'} = V_{\vec k,\vec k'} \braket{\phi_{c, \vec k}|\phi_{c,\vec k'}}
    \braket{\phi_{c, -\vec k}|\phi_{c,-\vec k'}}.
\end{equation}
At present there is no consensus on the microscopic pairing mechanism in rhombohedral multilayer graphene \cite{yang2025topological,bernevig2025berry,yoon2025quarter,gil2025charge,parra2025band,chou2025intravalley,gaggioli2025spontaneous,zhang2025pathways,han2025exact,geier2025chiral,qin2026chiral}, 
and identifying such a mechanism is not the goal of this work. Our focus is to determine the orbital magnetization of a given superconducting state. We therefore introduce a phenomenological attractive interaction that favors $p$-wave pairing,
\begin{equation}
\begin{split} \label{eq:V_pheno}
V_{\vec k,\vec k'} 
= -\frac{g}{A}  \left( e^{i\phi_{\vec k}} e^{-i\phi_{\vec k'}} + e^{-i\phi_{\vec k}} e^{i\phi_{\vec k'}} \right).
\end{split}
\end{equation}
Here $A$ denotes the system area, and we choose $g\sim160~\mathrm{meV\,nm^{2}}$, a value that yields a reasonable estimate of the superconducting transition temperature and its density dependence~\cite{chou2025intravalley}. 
In the case we study in the main text, the Fermi energy $\epsilon_F=0.92$ meV, $gn_e=0.8$ meV and $g N_F=0.5$ where $N_F$ is the density-of-states at the Fermi level. We note when $g$ is too small, superconductivity is no longer energetically favored and the system remains metallic despite the presence of an attractive interaction~\cite{chou2025intravalley}. This is because pairing in a time-reversal symmetry–broken band does not give rise to the conventional BCS logarithmic divergence. 

The mean-field approximation leads to the following 
BCS Hamiltonian for the conduction band, 
\begin{equation}
\begin{split} \label{eq:H_bcs_conduction}
&H_c^{\rm{BCS}}(\vec{k})
= \xi_{c,\vec k}^o- \vec d_\vec k \cdot \vec \tau,\\
&\vec d_\vec k = -[
\Re(\Delta_\vec k)\,,-\Im(\Delta_\vec k)\,,\xi_{c, \vec k}^e],
\end{split}
\end{equation}
where $\vec\tau$ denotes Pauli matrices in Nambu space.  $\xi_{c,\vec k}^{o} = (\xi_{c,\vec k}-\xi_{c,-\vec k})/2$ and $\xi_{c,\vec k}^{e} = (\xi_{c,\vec k}+\xi_{c,-\vec k})/2$ are the odd and even combination of band-dispersion. The superconducting gap $\Delta_{\vec k}=-\Delta_{-\vec k}$ is obtained self-consistently, as discussed in the Appendix, yielding the Bogoliubov quasiparticle energies $E_{\pm,c,\vec k}$ and eigenvectors $\lvert U_{\pm,c,\vec k}\rangle$, where $\pm$ label the electron-like and hole-like branches in the Nambu representation.

Fig.~\ref{fig:mf_solution}a shows the Bogoliubov quasiparticle band dispersion for tetralayer graphene ($N=4$)  with the inset indicating the electron and hole Fermi surfaces. Fig.~\ref{fig:mf_solution}b shows the phase winding of the $p-ip$ superconducting order parameter, which rotates by $2\pi$ upon traversing a closed loop counterclockwise in momentum space. Fig~\ref{fig:mf_solution}c shows the normalized Anderson pseudospin vector \cite{anderson1958random}, $\hat{\vec d}_{\vec k}=\vec d_{\vec k}/|\vec d_{\vec k}|$, revealing a Skyrmion-like texture. The out-of-plane pseudospin component $\hat d_{\vec k}^z$ encodes the joint occupation of states at $\vec k$ and $-\vec k$: $\hat d_{\vec k}^z=1$ indicates that both states are occupied, while $\hat d_{\vec k}^z=-1$ indicates that both are empty. In the normal state, $\hat d_{\vec k}^z$ exhibits a sharp domain wall at the Fermi surface, switching abruptly between $\pm 1$. In the superconducting state, pairing between $\vec k$ and $-\vec k$ smooths this domain wall. Because the normal-state Fermi surface has only $C_3$ symmetry, $\epsilon_{\vec k}=\epsilon_{C_3\vec k}$, the resulting $\hat d_{\vec k}^z$ has a $C_6$ symmetry.
Since the mapping $\vec k\rightarrow \hat d_{\vec k}$  wraps the unit sphere once, it has a skyrmion (Pontryagin) index $1$, which serves as the topological invariant of the chiral superconductor. In rhombohedral multilayer graphene, however, $\hat d_{\vec k}^z$ still varies sharply in momentum space near the overlap of the electron and hole Fermi surfaces. This behavior originates from the unusual normal-state dispersion, which is exceptionally flat near the $K$ point but rises rapidly at larger $|\vec k|$, producing the so-called “Berry-trashcan” profile~\cite{bernevig2025berry}. 
Although our phenomenological interaction $V$ [Eq.~\eqref{eq:V_pheno}] does not energetically distinguish between $p+ip$ over $p-ip$ pairing, the sublattice winding encoded in the form factors in $\tilde{V}$ selects a definite chirality \cite{geier2024chiral,may2025pairing}. The resulting energy splitting between the two chiral states, however, remains small; for the case shown in Fig.~\ref{fig:mf_solution}, it is approximately $44~\mu\mathrm{eV}$ per electron. We find that the energetically favored superconducting state has a chirality \emph{same} to that of the normal-state sublattice winding, see Appendix.
%the conduction band Berry curvature is negative,

%For valley $K$, where the intralayer hopping between sublattices is $\langle A_1|T|B_1\rangle=\hbar v_F(\tau k_x+ik_y)$ with $\tau=+1$, the energetically favored state has $p-ip$ chirality.
%This mean-field solution provides the basis for analyzing the orbital magnetization $M_z$.  In the following, we focus on the highest-density point shown in Fig.~\ref{fig:schematic}, whose superconducting solution is analyzed in detail in Fig.~\ref{fig:mf_solution}; results for the remaining density points are presented in Appendix.

%\begin{figure}[t]
%    \centering
%\includegraphics[width=0.8\linewidth]{schematic_v2.pdf}
%\caption{ 
%The NN contribution represents a background inherited from the normal state, whereas the superconductivity-induced orbital response arises from the NB, BN, and BB transitions.}
%\label{fig:transition_schematic}
%\end{figure}

\subsection{Normal-normal (NN) transitions}
%The normal-normal contribution is the simplest: it consists of transitions from all valence bands of $h^{K\cdot p}$ to all non-superconducting conduction bands $c'\neq c$.  In the general expression Eq.~\eqref{eq:M_general}, this corresponds to choosing initial state $i=(v,+)$ and final state $m=(c',+)$, with $v$ running over all valence bands and $c'\neq c$ running over all conduction bands except the one participating in superconductivity. An equivalent set of contributions arises from choosing $i=(c',-)$ and $m=(v,-)$, which describe the same virtual processes in the hole sector of the Nambu representation.Because these two descriptions are related by particle–hole conjugation, they yield identical contributions. Adding both, we obtaincorresponds to choosing initial state $i=(v,+)$ and final state $m=(c',+)$, with $v$ running over all valence bands and $c'\neq c$ running over all conduction bands except the one participating in superconductivity. An equivalent set of contributions arises from choosing $i=(c',-)$ and $m=(v,-)$, which describe the same virtual processes in the hole sector of the Nambu representation.Because these two descriptions are related by particle–hole conjugation, they yield identical contributions. Adding both, we obtain

\begin{figure*}[t]
    \centering
    \includegraphics[width=1.0\linewidth]{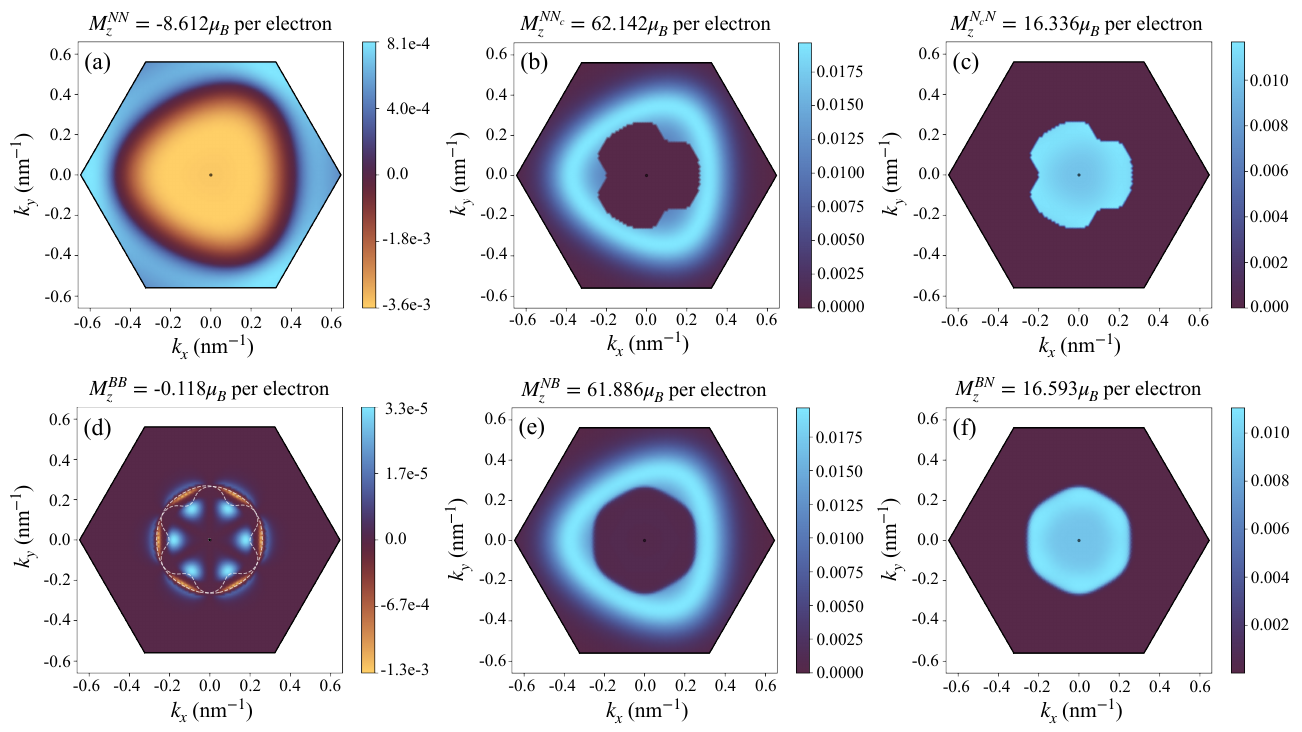}
\caption{
Momentum-space distributions of the orbital magnetization contributions from the normal--normal, mixed (NB and BN), and Bogoliubov--Bogoliubov channels:
(a) $M_z^{\rm NN}(\vec k)$,
(b) $M_z^{\rm NN_c}(\vec k)$,
(c) $M_z^{\rm N_cN}(\vec k)$,
(d) $M_z^{\rm BB}(\vec k)$,
(e) $M_z^{\rm NB}(\vec k)$, and
(f) $M_z^{\rm BN}(\vec k)$,
shown in units of $\mu_B$ per electron for rhombohedral tetralayer graphene in valley $K$ at $V_z=41$~meV and carrier density $n_e=0.5\times10^{12}$~cm$^{-2}$. 
The total orbital magnetization obtained by summing over $\vec k$ is indicated in the title of each panel. 
In panel (d), the color scale spans the minimum and maximum values of $M_z^{\rm BB}(\vec k)$; note that the peak magnitude is nearly two orders of magnitude larger than the background, so the dominant contribution originates from the region where the electron and hole Fermi surfaces overlap, shown as the dashed white contours. 
By contrast, the mixed NB and BN contributions arise from the entire Fermi sea rather than being confined to the Fermi-surface region.
}

\label{fig:allMz}
\end{figure*}

The normal–normal sector corresponds to choosing $(m,i)$ in the general expression Eq.~\eqref{eq:M_general} to be the normal bands only. It describes a background orbital magnetization inherited from the parent normal state.  It vanishes when the parent state is time-reversal symmetric. For rhombohedral multilayer graphene, the parent state is a quarter-metal that breaks time-reversal symmetry so this contribution is finite and it take the form,
\begin{align} 
\begin{split}
M_z^{\rm NN}&(\vec k) =\frac{e}{2\hbar} \sum_{c'\neq c}\sum_{v} \frac{\xi_{c' \vec k} + \xi_{v \vec k}}{(\xi_{c' \vec k} - \xi_{v \vec k})^2} \epsilon_{\mu \nu}\times  \\ &\mathrm{Im} \left( \bra{\phi_{c'\vec k}} (\partial_\nu h(\vec k))\ket{\phi_{v\vec k}} \langle \phi_{v \vec k} | (\partial_\mu h(\vec k))| \phi_{c' \vec k} \rangle \right). 
\end{split}
\end{align}
Here $v$ runs over all valence bands and $c'$ runs over all conduction bands except the one participating in superconductivity ($c$). The k-space distribution is shown in Fig.~\ref{fig:allMz}a, the large contribution comes from the zone corner ($k_x=k_y=0$) because the Berry curvature is strongest over there.

\subsection{Mixed Normal–Bogoliubov Transitions (NB and BN)}The next class of processes corresponds to transitions in which one of the indices $(m,i)$ is a Bogoliubov band, giving rise to the mixed normal–Bogoliubov term,  $M^{\rm NB}_z$ and $M^{\rm BN}_z$:
\begin{align} 
&M_z^{\rm NB}(\vec k) =\frac{e}{2\hbar}
\sum_{v} \bigg[\frac{E_{c,+,\vec k} + \xi_{v \vec k}}{(E_{c,+,\vec k} - \xi_{v \vec k})^2} \sin^2\left(\frac{\theta_\vec k}{2}\right)\epsilon_{\mu \nu} \nonumber \\  &\mathrm{Im} \left( \bra{\phi_{c\vec k}} (\partial_\nu h(\vec k))\ket{\phi_{v\vec k}} \langle \phi_{v \vec k} | (\partial_\mu h(\vec k))| \phi_{c \vec k} \rangle \right) \bigg]. \\
&M_z^{\rm BN}(\vec k) =\frac{e}{2\hbar}
\sum_{c'\neq c} \bigg[ \frac{E_{c,-,\vec k} + \xi_{c' \vec k}}{(E_{c,-,\vec k} - \xi_{c' \vec k})^2} \cos^2\left(\frac{\theta_\vec k}{2}\right)\epsilon_{\mu \nu}   \nonumber
\\ &\mathrm{Im} \left( \bra{\phi_{c'\vec k}} (\partial_\nu h(\vec k))\ket{\phi_{c\vec k}} \langle \phi_{c \vec k} | (\partial_\mu h(\vec k))| \phi_{c' \vec k} \rangle \right)\bigg]. 
\end{align}

The term $M_z^{\rm NB}$ is obtained by fixing the unoccupied state $m$ to be the upper Bogoliubov band and summing the occupied state $i$ over all occupied normal-sector BdG states. Accordingly, the label $v$ in Eq.~(25) includes not only the electron branches of filled normal-state valence bands, but also the occupied hole-conjugate branches of empty normal-state conduction bands. Conversely, $M_z^{\rm BN}$ is obtained by fixing the occupied state $i$ to be the lower Bogoliubov band and summing the unoccupied state $m$ over all unoccupied normal-sector BdG states. Thus, the label $c'$ in Eq.~(26) includes both the electron branches of empty normal-state conduction bands and the unoccupied hole-conjugate branches of filled normal-state valence bands. Importantly, the matrix elements appearing in $\mathrm{Im}(\cdots)$ are entirely determined by normal-state wavefunctions, since superconducting pairing is restricted to the first conduction band. Beyond the replacement of the normal-state dispersion $\xi_{c\vec k}$ by the Bogoliubov quasiparticle energies $E_{c,\pm,\vec k}$, the most notable feature of the above Equations is the appearance of the BCS probability factors $\cos^2(\theta_{\vec k}/2)$ and $\sin^2(\theta_{\vec k}/2)$. These factors represent the probabilities that the states $\vec k$ and $-\vec k$ are occupied or emptied, respectively. We emphasis that this modification is a Fermi-sea effect and is not confined to the vicinity of the Fermi surface.

To obtain a quantitative understanding, it is informative to ask how the orbital magnetization behaves when we take the normal limit. This can be done by replacing the BCS coherence factors and quasiparticle energies according to
$\cos^2\left(\theta_{\vec k}/2\right) \rightarrow \Theta(\xi_{c\vec k})$, $
E_{c,\pm,\vec k} \rightarrow \xi_{c\vec k}$
where $\Theta(x)$ is the Heaviside step function. The resulting normal-state expressions are denoted by $M^{\rm NN_c}$ and $M^{\rm N_cN}$. These contributions are shown in Fig.~\ref{fig:allMz}.
In the normal state, $M_z^{\rm N_cN}(\vec k)$ is nonzero only for momenta occupied in the first conduction band. In the superconducting phase, this strict Pauli constraint is relaxed by the BCS probability factors, allowing contributions from a broader range of momenta and thereby enhancing the orbital magnetization, c.f.~$M_z^{\rm BN}(\vec k)$. By contrast, $M_z^{\rm NN_c}(\vec k)$ is finite only for momenta unoccupied in the first conduction band, and the corresponding Bogoliubov term $M_z^{\rm BN}(\vec k)$ suppresses this contribution.
For the single--Fermi-surface case considered here, the mixed normal--Bogoliubov terms have a negligible net effect on the orbital magnetization,
$M_{z}^{\rm BN} + M_{z}^{\rm NB} \simeq M_z^{\rm N_cN} + M_{z}^{\rm NN_c}$.
The dominant modification instead arises from the pure Cooper-pair contribution, which we discuss next.

\subsection{Bogoliubov-Bogoliubov (BB) transitions}

The Bogoliubov-Bogoliubov (BB) transitions describe the interband processes that occur entirely within the superconducting quasiparticle sector. It is obtained by setting $(m,i)$ to be upper and lower Bogoliubov band in Eq.~\eqref{eq:M_general}, which yields
\begin{equation} 
\begin{split}\label{eq:M_BB}
M_z^{\rm BB}(\vec k) &= 
\frac{e\hbar}{8} 
\left(\frac{\epsilon_{c,\vec k}-\epsilon_{c,-\vec k}}{2}\right)
\frac{\mathcal{V}^I_\vec{k} \times \gamma^R_\vec{k}+
\mathcal{V}^R_\vec{k} \times 
\gamma^I_\vec{k}  }{(\xi_{c,\vec k}^{e})^2 + |\Delta_{\vec k}|^2}.
\end{split}
\end{equation}
Here 
$\hbar^{-1}\langle U_{c,- ,  \vec k} | \nabla H^{\rm BCS}_c(\vec k))| U_{c,+ , \vec k} \rangle \equiv   \mathcal{V}^R_\vec{k}  +i\mathcal{V}^I_\vec{k}$ are the quasiparticle velocity matrix elements 
 and $ \bra{U_{c,+ ,\vec k}} \vec{\Gamma}_{\vec k}\ket{U_{c,- , \vec k}} 
\equiv\vec{\gamma}^R_\vec{k}  +i\vec{\gamma}^I_\vec{k} $ are the dressed photon vertex matrix elements. 
A correct description of the orbital magnetization carried by Cooper pairs must involved this dressed photon vertex in order to preserve gauge invariance. Determining $\vec{\Gamma}_{\vec k}$ is equivalent to solving the linearized time-dependent Bogoliubov–de Gennes equations (diagonalizing the associated stability matrix). We describe the calculation of $\vec{\Gamma}_{\vec k}$ in the next section, and present the resulting $M^{\rm BB}_z$ in Fig.~\ref{fig:allMz}(d). Unlike the mixed NB and BN contributions, where superconductivity enters only through quasiparticle energies and occupation factors, the BB contribution depends explicitly on the chirality of the superconducting order parameter. While the $M^{\rm NB}_z$ and $M_z^{\rm BN}$ come from the entire Fermi sea,  $M_z^{\rm BB}$ is strongly peaked around the overlap region of the electron and hole Fermi surfaces. The latter is consistent with the physical intuition that the intrinsic orbital magnetization of Cooper pairs is dominated by states near the Fermi surface.

\begin{figure*}[t]
    \centering
    \includegraphics[width=1.0\linewidth]{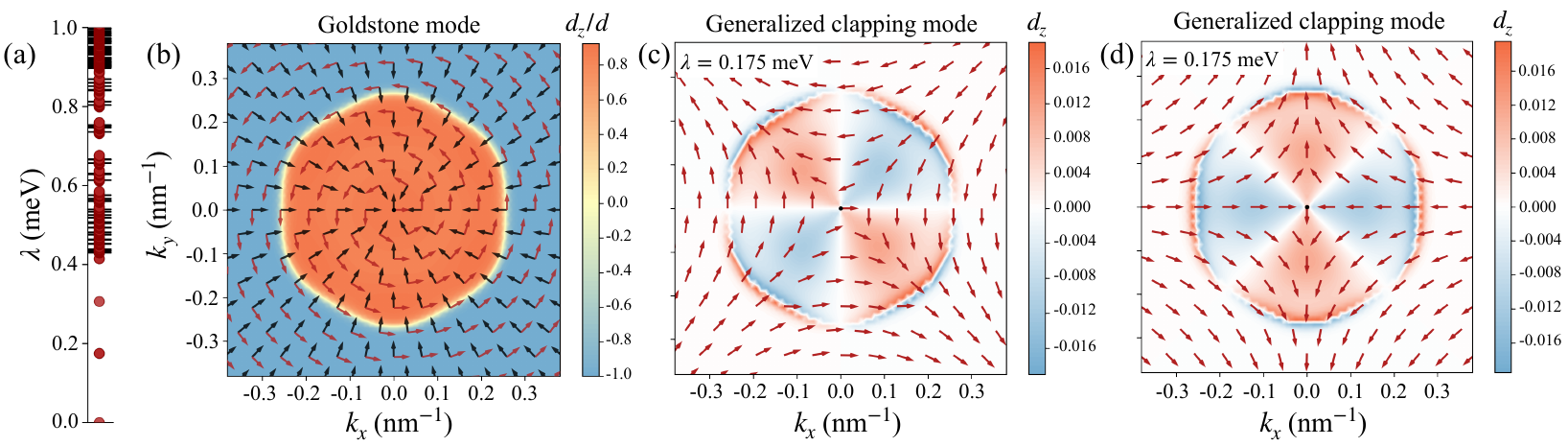}
    \caption{(a) Eigenvalues $\lambda$ of the stability matrix $\mathcal{S}$ (red dots) and $\Delta E_{\bf k}$ (black dash) in Eq.~\eqref{eq:S_matrix} for rhombohedral tetralayer graphene in valley $K$ at $V_z=41$~meV and carrier density $n_e=0.5\times10^{12}$~cm$^{-2}$. (b) The Anderson-Bogoliubov-Goldstone mode ($\delta \vec{d}^{\rm ABG}_\vec k$, red arrows) superimposed on top of the groundstate pseudospin distribution (black arrows). (c-d) The pseudospin distribution of the two-fold generalized clapping modes ($\delta \vec{d}^{\rm CGM}_\vec k$). They have opposite chirality ($p+ip$) compared to that of the ground state ($p-ip$). 
    %In (b-d), the in-plane components $(d_x,d_y)$ are normalized by $(d_x^2+d_y^2)^{1/2}$.
    %, and $d = (d_x^2+d_y^2+d_z^2)^{1/2}$ in (b).
    }
\label{fig:collective_modes}
\end{figure*}

Another important feature of Eq.~\eqref{eq:M_BB} is that it vanishes in the presence of momentum-negation symmetry (spinless time-reversal or inversion symmetry), $\epsilon_{c,-\vec k}=\epsilon_{c,\vec k}$. Consequently, symmetry forbids any intrinsic orbital magnetization generated by zero-center-of-mass momentum Cooper pairs when the parent band dispersions has even-fold rotational symmetry $C_{2n}$ 
$(n=0,1,2...)$, such as a circular Fermi surface.
In fact, as we show in the next section, not only does the energy prefactor in Eq.~\eqref{eq:M_BB} vanish when $\epsilon_{c,-\vec k}=\epsilon_{c,\vec k}$, but the photon-vertex matrix elements also vanish. 
%As a result, $M_z^{\rm BB}$ (for zero center of mass pairing) is strongly suppressed as the Fermi surface approaches a circular shape.

%Moreover, while the NB and BN terms receive contributions from the entire Fermi sea, $M^{BB}_z$ is predominantly concentrated in the momentum region where the electron and hole Fermi surfaces overlap, as shown by the dashed white contours in Fig.~\ref{fig:allMz}d. 
%This is consistent with the physical origin of $M^{BB}_z$, which arises primarily from Cooper pairing processes that are most effective near the Fermi surface. 
%Here $H^{\rm BCS}_c(\vec k)$ is the conduction-band BCS Hamiltonian introduced in Eq.~\eqref{eq:H_bcs_conduction}, and  $|U_{c,\pm,\vec k}\rangle$ and $E_{c,\pm,\vec k}$ are its eigenstates and eigenvalues. The only undetermined quantity in Eq.~\eqref{eq:M_BB} is the dressed photon vertex $\vec{\Gamma}_{\vec k}$.  

\section{Results}
\subsection{Generalized Clapping Mode}
To determine the contribution of Cooper pairs to the orbital magnetization, we must compute the dressed photon vertex $\vec\Gamma_{\vec k}$. This renormalization mainly comes from the collective excitations of the superconducting order parameter. We begin by simplifying Eq.~\eqref{eq:X_linearized}, which can be recast in the compact matrix form
\begin{equation} \label{eq:S_matrix}
    \sum_{\mathbf{k}'}
\begin{pmatrix}
\Delta E_{\mathbf{k}}\,\delta_{\mathbf{k}\mathbf{k}'} + C^{RR}_{\mathbf{k},\mathbf{k}'} &
C^{RI}_{\mathbf{k},\mathbf{k}'}\\[4pt]
C^{IR}_{\mathbf{k},\mathbf{k}'} &
\Delta E_{\mathbf{k}}\,\delta_{\mathbf{k}\mathbf{k}'} + C^{II}_{\mathbf{k},\mathbf{k}'}
\end{pmatrix}
\begin{pmatrix}
X^R_{\mathbf{k}'}\\[2pt]
X^I_{\mathbf{k}'}
\end{pmatrix}
=
\begin{pmatrix}
-e\vec{\gamma}^{\,0}_{\mathbf{k}}\cdot\vec{A}\\[2pt]
0
\end{pmatrix}.
\end{equation}
Here $\Delta E_{\vec k}=E_{c,+,\vec k}-E_{c,-,\vec k}$ is the $k$-local energy difference between the two Bogoliubov band and
$\langle U_{c,+\vec k}|\delta P|U_{c-,\vec k}\rangle =X_{\vec k}^R + i X_{\vec k}^I$ 
is the response amplitude. $C$ describes the interactions between pseudospin-flips.
The matrix on the right is termed the stability matrix $\mathcal{S}$ and the left hand side is the bare paramagnetic response $
   \vec\gamma^{\,0}_{\vec k}
=
\sin\theta_{\vec k}[\vec v_{cc}(\vec k)+\vec v_{cc}(-\vec k)]/2
$.  This method of studying response is essentially a linearized spin-wave calculation applied to pseudospins in momentum space, following Ref.~\cite{anderson1958random}.
%See the Appendix for further discussion.

%For a given pseudospin direction $\hat d_{\vec k}$, the unit vectors $\vec e_1(\vec k)$ and $\vec e_2(\vec k)$ span the tangent plane of the Bloch sphere. These vectors provide a local basis for transverse pseudospin fluctuations. In superconductors, the exchange field of the Anderson pseudospin is given by the pairing self-energy, which lies entirely in the particle–hole mixing (xy) plane.
%As a result, unlike in real ferromagnets where the exchange interaction favors a
% alignment of transverse fluctuation directions across momentum space, the
%superconducting self energy interaction is sensitive only to the components of these
%fluctuations projected into the xy plane. The vectors $\vec a_{\vec k}$ and $\vec b_{\vec k}$ are nothing but the projections of thelocal transverse fluctuation directions onto the particle–hole mixing plane, $\vec a_{\vec k}=P_{xy}\vec e_1(\vec k)$ and $\vec b_{\vec k}=P_{xy}\vec e_2(\vec k)$, where $P_{xy}=\text{diag(1,1,0)}$. When these projected components are aligned (orthogonal), they interact through $\tilde V^R$  ($\tilde V^I$).

In the language of time-dependent mean-field, the $C$ matrix element describes the scattering between a pair of zero–center-of-mass-momentum Bogoliubov quasiparticles $\gamma^\dagger_{\vec k}\gamma^\dagger_{-\vec k}|0\rangle$, and another 
pair
$\gamma^\dagger_{\vec k'}\gamma^\dagger_{-\vec k'}|0\rangle$.
The eigenvalue spectrum of $\mathcal S$ is shown in Fig.~\ref{fig:collective_modes}a. 
The black dashed line corresponds to the energy of two uncoupled Bogoliubov quasiparticles propagating independently, obtained by setting $C=0$. In this limit no collective modes are present and the spectrum consists of a continuum starting at $2\Delta$. The red points show the spectrum for $C\neq0$, where the most prominent feature is the emergence of discrete collective modes below the continuum.  We can visualize the  wavefunction of the collective mode through the associated pseudospin distribution,
$
\delta\vec d_{\vec k}
=
2\big[
\delta F^R_{\vec k},
-\delta F^I_{\vec k},
X_{\vec k}^R\sin\theta_{\vec k}
\big],$ where the $\delta F_{\vec k}^{R/I}$ real and imaginary components of the pairing-amplitude fluctuation.

%are given by  $ \delta F_{\vec k}^R=X_{\vec k}^R a^x_ \vec k  + X_{\vec k}^I b^x_\vec k  $ and $ \delta F_{\vec k}^I = -X_{\vec k}^R a^y_{\vec k} - X_{\vec k}^I b^y_{\vec k}$.

The zero-energy mode corresponds to the Anderson–Bogoliubov–Goldstone (ABG) mode. 
In Fig.~\ref{fig:collective_modes}b, we plot the corresponding pseudospin texture $\delta\vec d_{\vec k}^{\rm ABG}$ (red arrows) superimposed on the ground-state texture $\vec d_{\vec k}$ (black arrows). The ABG mode represents a global rotation of $\vec d_{\vec k}$, such that $\delta\vec d_{\vec k}^{\rm ABG}$ is locally orthogonal to the ground-state pseudospin configuration.

Fig.~\ref{fig:collective_modes}c and d show a pair of closely-degenerate collective modes at energy $\lambda=0.175~\mathrm{meV}$, which are the generalized clapping mode (GCM) \cite{Levitan_RMG_2025, Wolfle_collective_1976, Hirashima1988, Hirschfeld_EM_1989, Hirschfeld_collective_1992, Yip1992, Kee_collective_2000, Higashitani_2000, Higashitani_EM_2000, Balatsky_2000,sauls2015anisotropy}. Their pseudospin texture, $\delta\vec d_{\vec k}^{\rm GCM}$, corresponds to a $p$-wave order parameter with winding opposite to that of the ground state $\vec d_{\vec k}$. 
The GCM describes collective fluctuations from the energetically favored chirality to the higher-energy, unfavored chirality. Because the two chiralities correspond to an Ising-like 
$Z_2$ degree of freedom, this mode is generically gapped but remains below the quasiparticle continuum.
We find that the gap of the mode is reduced by the sublattice winding form factors by approximately a factor of two, which originates from the same quantum-geometric factors that contribute to the superfluid stiffness \cite{peotta2015superfluidity,peotta2023quantum}. The pseudospin textures of the two nearly degenerate GCM modes have orientations perpendicular to each other; i.e., one mode is the Goldstone partner of the other. Importantly, these modes are charge neutral, $\sum_{\vec k} \delta\vec d_{\vec k}^{\rm GCM} \cdot \hat z = 0 .$, and therefore do not hybridize with plasmons via the Anderson–Higgs mechanism. They can thus be probed optically independently from plasma excitations.

To compute the vertex function, we substitute
$X_{\vec k} = -\,e\,\vec\gamma_{\vec k}\cdot\vec A/\Delta E_{\vec k}
$ into Eq.~\eqref{eq:X_linearized} which yield,
\begin{equation}\label{eq:vertex}
    \sum_{\mathbf{k}'}
\begin{pmatrix}
 \delta_{\mathbf{k}\mathbf{k}'} + C^{RR}_{\mathbf{k},\mathbf{k}'}/\Delta E_{\vec k'} &
C^{RI}_{\mathbf{k},\mathbf{k}'}/\Delta E_{\vec k'}\\[4pt]
C^{IR}_{\mathbf{k},\mathbf{k}'}/\Delta E_{\vec k'} &
\delta_{\mathbf{k}\mathbf{k}'} + C^{II}_{\mathbf{k},\mathbf{k}'}/\Delta E_{\vec k'}
\end{pmatrix}
\begin{pmatrix}
\gamma^R_{\mathbf{k}'}\\[2pt]
\gamma^I_{\mathbf{k}'}
\end{pmatrix}
=
\begin{pmatrix}
\vec{\gamma}^{\,0}_{\mathbf{k}}\\
0
\end{pmatrix}.
\end{equation}
The dressed photon vertex $\vec\gamma_{\vec k}$ is obtained by inverting this matrix equation. In performing this matrix inversion, the ABG  mode must be removed, as it represents a spurious (unphysical) excitation. 
This is analogous to collective-motion calculations in nuclear physics~\cite{thouless2013quantum,rowe2010nuclear}, where zero-energy modes corresponding to rigid translation or rotation of the nucleus must be excluded from the physical response.

\subsection{Paramagnetic Response, Orbital Magnetization, and the Meissner Effect}

\begin{table}[t] \centering \begin{tabular}{c|c|c}
\hline & 
\begin{tabular}[c]{@{}c@{}}
2D chiral\\
SC in RMG
\end{tabular}
 & \begin{tabular}[c]{@{}c@{}}
3D bulk\\
conventional SC
\end{tabular}
\\ \hline TRS at $H=0$ & broken & preserved \\ 
\hline
Spontaneous $M$ & $M \neq 0$ & $M =0$ \\ 
\hline
Field-induced  $M_{\rm ind}$ & --  & 
\begin{tabular}[c]{@{}c@{}@{}} $M_{\rm ind}=-H/4\pi$, \, $H<H_{c1}$ 
\\ vortex lattice, \, $H_{c1}<H<H_{c2}$ 
\end{tabular} \\ 
\hline \end{tabular} 
\caption{
Spontaneous orbital magnetization $M$ versus magnetization $M_{\rm ind}$ induced by an external field $H$.
We focus on computing $M$ for 2D chiral superconductors in rhombohedral multilayer graphene (RMG). The parent state is a valley-imbalanced quarter metal that breaks time-reversal symmetry (TRS) and already has a finite spontaneous orbital magnetization.
}
\label{table:Meissner}
\end{table}

The optical activity of the GCM, in contrast to the optically dark response of a conventional BCS superconductor, and the finite Cooper-pair orbital magnetization share a common microscopic origin: nonvanishing matrix elements of the paramagnetic coupling. We now explain this connection in greater detail.

Eq.~\eqref{eq:S_matrix} describes the paramagnetic response of the superconducting state to a vector potential, while Eq.~\eqref{eq:vertex} determines the corresponding dressed photon vertex.
In both cases, the source term is the same bare vertex appearing on the right-hand side,
$\vec{\gamma}_{\mathbf{k}}^{\,0}
\equiv \langle U_{c,+,\mathbf{k}} | \Gamma^{0} | U_{c,-,\mathbf{k}} \rangle=\sin\theta_{\vec k}\big[\vec v_{cc}(\vec k)+\vec v_{cc}(-\vec k)\big]/2
$, which
corresponds to an Anderson pseudospin-flip matrix element.
This matrix element vanishes when momentum negation symmetry,
$\epsilon_{c,\mathbf{k}} = \epsilon_{c,-\mathbf{k}}$,
is present (either time-reversal or inversion symmetry),
and consequently the orbital magnetization $M_z^{\rm BB}$ vanishes. An Anderson pseudospin flip corresponds to the creation of a pair of Bogoliubov quasiparticles with zero center-of-mass momentum \cite{schrieffer2018theory}.

This cancellation is closely related to the microscopic origin of the Meissner effect and the reason why conventional BCS superconductor is optically dark ($q=0$).
At its core, the Meissner effect arises because the paramagnetic current response does not contribute in the static $\omega=0$, long-wavelength limit $q \to 0$, while the diamagnetic term remains finite and produces perfect diamagnetism.
Microscopically, this happens because the probability amplitude for creating two Bogoliubov quasiparticles from the BCS ground state vanishes at $q=0$, due to 
microscopic quantum interference effects in the BCS theory of superconductivity \cite{cooper1972microscopic,schrieffer2018theory}:
\begin{align}
    \vec{v_k}\cdot \vec{A}c^{\dagger}_\vec{k}c_\vec{k}|\textrm{BCS}\rangle &= 
    \vec{v_k}\cdot \vec{A} \;(u_\vec k v_\vec k) \gamma_\vec{k}^\dagger \gamma_{-\vec k}^\dagger|\textrm{BCS}\rangle \\
    \vec{v_{-k}}\cdot \vec{A}c^{\dagger}_{-\vec{k}}c_{\vec{-k}}|\textrm{BCS}\rangle &= 
    \vec{v_{-k}}\cdot \vec{A} \;(u_\vec k v_\vec k) \gamma_\vec{k}^\dagger \gamma_{-\vec k}^\dagger|\textrm{BCS}\rangle
\end{align}
The sum of the two contributions vanishes when the two members of the Cooper pair have the same energy $\vec{v_k}+\vec{v_{-k}}=\nabla_\vec k(\epsilon_\vec k+\epsilon_{-\vec k})=0$. This cancellation removes the paramagnetic contribution in the static long-wavelength limit, leaving the diamagnetic term responsible for perfect screening.

Since only the $q=0$ component of the paramagnetic response to the vector potential $\vec{A}$ enters the orbital magnetization
[c.f.~Eq~\eqref{eq:X_matrix}],
the orbital magnetization likewise vanishes in a conventional superconductor as a consequence of superconducting coherence factors.
Thus, a finite superconducting orbital magnetization $M_{BB}$ requires a nonvanishing zero-momentum pseudospin-flip vertex, which in the present problem is enabled by the broken time-reversal symmetry of the valley-polarized normal state. This spontaneous magnetization should be distinguished from field-induced Meissner screening. The present theory computes only the spontaneous orbital magnetization at $H=0$, as summarized in Table~\ref{table:Meissner}. Extending the theory to finite applied field, where spontaneous currents coexist with field-induced screening currents and possible vortex configurations, is left for future work.

Finally, detecting the sub-THz generalized clapping mode, which lies below the quasiparticle continuum at approximately $0.175$ meV for the parameters considered here, would provide a characteristic signature of chiral superconductivity in RMG.

\subsection{Orbital Magnetization Across the Lifshitz Transition}
For the single Fermi-surface we discussed so far, we found that orbital magnetization suppressed the quarter metal orbital magnetization by approximately $0.118\,\mu_B$ per electron, and they are coming from
intrinsic contribution from the Cooper pairs themselves, $M_z^{\rm BB}$.  This behavior, however, is not generic. At lower carrier densities, the single Fermi surface undergoes a Lifshitz transition and splits into three pockets. While the ground state in this regime remains a chiral $p-ip$ superconductor, it is no longer topological \cite{chou2025intravalley}. In this multi-pocket regime, $M^{\rm BB}_z$ becomes negligible, and the resulting change in orbital magnetization is dominated by the mixed NB and BN contributions, which enhance the orbital magnetization relative to the parent state, as shown in Table~\ref{tab:Mz_maintex}.

Overall, we find that superconductivity does not strongly modify the orbital magnetization of its the parent state, which is a valley-imbalanced orbital ferromagnetic metal.  This is important because the parent state
already possesses a large orbital magnetization compared to competing phases such as intervalley-coherent states. Consequently, the chiral superconducting state retains a large orbital magnetization relative to other competing states, allowing it to gain magnetic-field energy and potentially win the energetic competition in an external magnetic field. This result provides a way to comprehend the striking experimental observations of superconductivity stabilized by (orbital) magnetic field.

\begin{table}[t]
    \centering
    \begin{tabular}{c|c|c|c|c|c}
    \hline
          & $M_z^{\rm NN}$ & $M_z^{\rm NN_c} + M_z^{\rm N_cN}$ & $M_z^{\rm NB} + M_z^{\rm BN}$ & $M_z^{\rm BB}$ & $M_z^{\rm S}$ - $M_z^{\rm QM}$ \\ \hline
         1 & -18.2039  & 207.9884 & 208.4131 & -0.0095 & 0.4152 \\ \hline
         2 & -12.0827 & 138.4756 & 138.4896 & 0.0051 & 0.0191 \\ \hline
         3 & -9.4704 & 102.4586 & 102.4170 & -0.0070 & -0.0486 \\ \hline
         4 & -8.8083 & 89.8895 & 89.8505 & -0.0232 & -0.0622 \\ \hline
         5 & -8.6118 & 78.4780 & 78.4787 & -0.1184 & -0.1177 \\ \hline
    \end{tabular}
    \caption{ Different contributions to $M_z$ (in unit of $\mu_B$ per electron) for the five data points labeled 1 to 5 from left to right in Fig.~\ref{fig:schematic}(b). $M_z^{\rm QM}=M_z^{\rm NN}+M_z^{\rm NN_c} + M_z^{\rm N_cN}$ and
    $M_z^{\rm S}=M_z^{\rm NN}+M_z^{\rm NB} + M_z^{\rm BN}+M_z^{\rm BB}$.}
    \label{tab:Mz_maintex}
\end{table}

\section{Discussion and Outlook}
We have developed a microscopic theory for computing orbital magnetization in chiral superconductors. The key main distinction between the orbital magnetization in the normal and superconducting state is that the quasiparticle velocity operator and the photon vertex become distinct. We show how orbital magnetization can be understood in terms of different classes of transitions (NN, NB, BB), thereby resolving a long-standing conceptual challenge in defining magnetization in superconducting systems. Normal–Bogoliubov processes involve the Fermi sea, while Bogoliubov–Bogoliubov processes are concentrated at the Fermi surface. Our formalism is general and extends to three-dimensional systems and to  moiré minibands in twisted semiconductors ~\cite{xu2025signatures,guerci2025fractionalization,wang2025chiral,jahin2026enhanced}.

For rhombohedral multilayer graphene, the theory yields several experimentally testable predictions. The superconductivity-induced change in orbital magnetization should change sign across Lifshitz transitions and correlate with Fermi-surface areas measured via quantum oscillations. The Cooper-pair contribution is controlled by Fermi-surface anisotropy and vanishes in the circular limit. We further predict a pair of closely degenerate, gapped generalized clapping modes that are optically active in the sub-THz regime.

Our theory can be readily extended to inhomogeneous and finite-momentum pairing states and to strong-field limit using magnetic Bloch states \cite{norman1994magnetic,macdonald1993quantum,franz2000quasiparticles,yasui2002theory,murray2015majorana}. These extensions will further broaden the scope and applicability of our formulation.

We emphasize that our work computes the spontaneous orbital magnetization from the $q\rightarrow 0$ response of a translationally invariant system to a slowly varying magnetic field. This bulk quantity is directly relevant to thermodynamic probes, such as the slope of the phase boundary between the chiral superconducting and normal metallic states in the $n$–$B$ plane. The relationship between the $q\rightarrow 0$ response and a calculation directly in a uniform magnetic field is discussed in Ref.~\cite{huang2026orbital} by one of us. NanoSQUID measurements, by contrast, detect the stray magnetic flux generated by the spontaneous magnetization and associated current distribution in a finite sample. A quantitative comparison with nanoSQUID images therefore requires accounting for the sample geometry, scan height and other parameters.
During the completion of this manuscript, we became aware of recent nanoSQUID measurements showing that the quarter-metal state retains a finite magnetic signal upon cooling below the superconducting transition temperature $T_c$, although the signal is somewhat weaker than in the normal state \cite{Sheekey_nanoSQUID_2026,Dutta_nanoSQUID_2026}.

\textbf{Acknowledgment:}
We acknowledge helpful discussions with Etienne Lantagne-Hurtubise, Liang Fu, Benjamin Levantin, Leo Li, James Sauls, Manfred Sigrist, Nemin Wei, Cong Xiao and Di Xiao. C.~Huang also thanks Giovanni Vignale for drawing his attention to Rudolf Peierls's book \textit{Surprises in Theoretical Physics}. J.~Zhu is supported by the U.S. Department of Energy, Office of Basic Energy Sciences, under Contract No. DE-SC0025327. C.~Huang is supported by the U.S. Department of Energy, Office of Science, Office of Basic Energy Sciences, under Award Number DE-SC-0024346.\\

\textbf{Author Contributions Statement:} 
Both authors contributed equally to the conceptual development, calculations, and writing of the manuscript.\\

\textbf{Competing Interests Statement:}
The authors declare no competing interests.
\\

\textbf{RESOURCE AVAILABILITY}\\

\textbf{Lead contact}
Requests for further information and resources should be directed to and will
be fulfilled by the lead contact,Chunli Huang (chunli.huang@uky.edu).

\textbf{Materials availability}
Materials availability
This study did not generate new materials.

\textbf{Data and code availability}
All simulation data are provided in a the paper.
All codes utilized in this study are third-party packages. Any additional information required to reanalyze the data reported in this paper is available from the lead contact upon request.

\bibliographystyle{apsrev4-2}
\bibliography{references}

%\end{document}
\newpage
\clearpage

\appendix

% Reset equation and figure counters
\setcounter{equation}{0}
\setcounter{figure}{0}

% Redefine numbering style
\renewcommand{\theequation}{A\arabic{equation}}
\renewcommand{\thefigure}{A\arabic{figure}}

\section{$K\cdot p$ Hamiltonian}
\begin{figure}[t]
    \centering
    \includegraphics[width=1.0\linewidth]{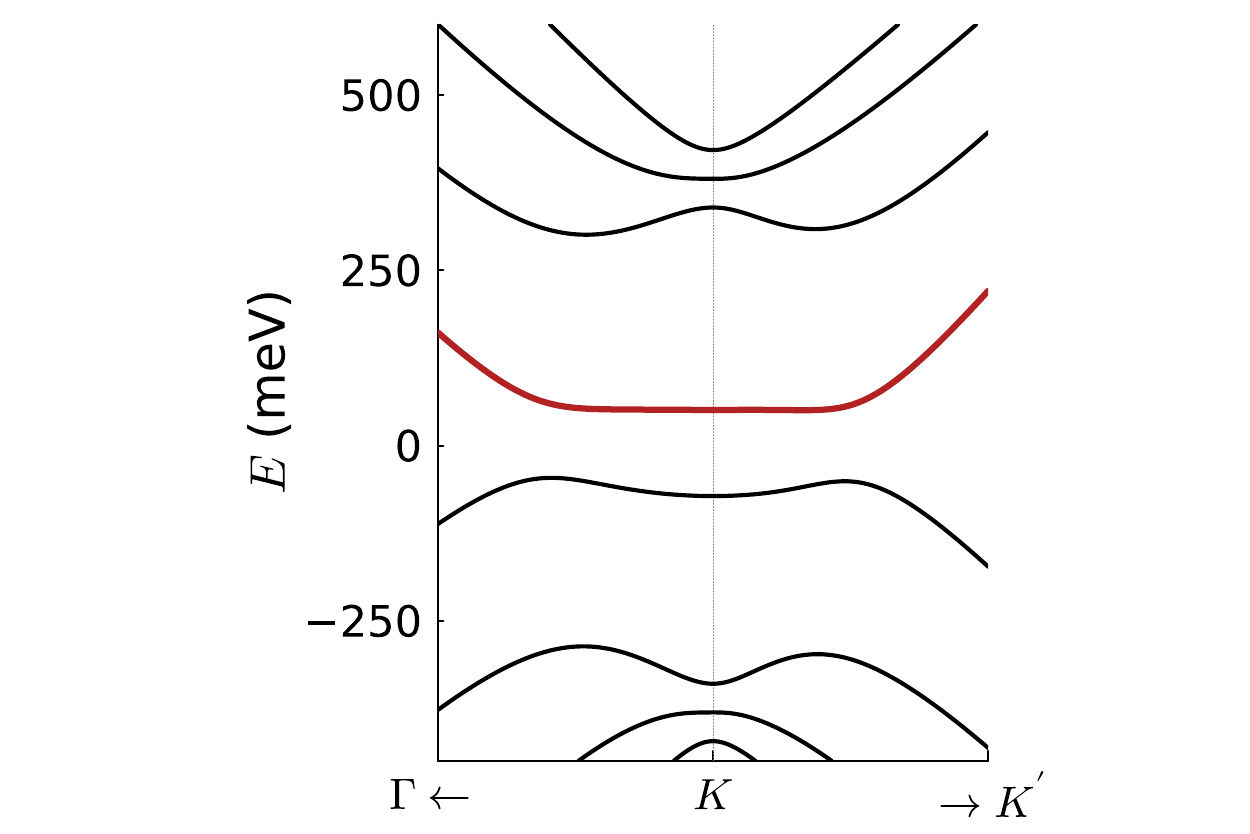}
    \caption{Single-particle band structure with $V_z=41$ meV. Superconductivity occurs in the lowest conduction band (red curve).} 
\label{figS:spectrum}
\end{figure}

The tetralayer graphene $K\cdot p$ Hamiltonian in the spinor basis $(A_1, B_1, A_2, B_2, A_3, B_3, A_4, B_4)$ is given by the following,
\begin{equation}
\begin{split}
&H_{\rm 4LG}(\vec{k}) = \\
&\begin{pmatrix}
h_{11}(\vec{k}) + \frac{3}{2} V_z & h_{12}(\vec{k}) & h_{13} & 0  \\
h^\dagger_{12}(\vec{k}) & h_{22}(\vec{k}) + \frac{1}{2} V_z & h_{12}(\vec{k}) & h_{13}  \\
h_{13} & h^\dagger_{12}(\vec{k}) & h_{33}(\vec{k}) - \frac{1}{2} V_z & h_{12}(\vec{k}) \\
0 & h_{13} & h^\dagger_{12}(\vec{k}) & h_{44}(\vec{k}) - \frac{3}{2} V_z 
\end{pmatrix}
\end{split}
\end{equation}
where
\begin{equation}
\begin{split}
&h_{ll}(\vec k) = 
\begin{pmatrix}
u_{A_l} & v_0 \vec{\pi}^\dagger \\
v_0 \vec{\pi} & u_{B_l}
\end{pmatrix},\\
&h_{12}(\vec k) = 
\begin{pmatrix}
-v_4 \vec{\pi}^\dagger & -v_3 \vec{\pi} \\
t_1 & -v_4 \vec{\pi}^\dagger
\end{pmatrix},\\
&h_{13}(\vec k) = 
\begin{pmatrix}
0 & t_2 \\
0 & 0
\end{pmatrix}.
\end{split}
\end{equation}
$\pi_\tau(\vec{k}) = \tau k_x + ik_y$, $\vec k$ is measured from valley $\vec{K}_\tau = (\tau, 0)4\pi/\sqrt{3}a$.
$v_j = \sqrt{3}a|t_j|/2$, $a=0.246$ nm is the lattice constant of graphene. In the calculations, we have used the model parameters adapted for rhombohedral trilayer graphene from Refs.~\cite{zhou2021half, chou2025intravalley}: $t_0 = -3100$ meV, $t_1 = 380$ meV, $t_2 = -7.5$ meV, $t_3 = 290$ meV, $t_4 = 141$ meV, $u_{A_1} = u_{B_4} = -10.5$ meV, $u_{B_1} = u_{A_2} = u_{B_2} = u_{A_3} = u_{B_3} = u_{A_4} = 0$. The band structure at interlayer potential $V_z=41$ meV is shown in Fig.~\ref{figS:spectrum}.

Ignoring the trigonal warping terms and the next-layer tunneling, $v_3=v_4=0$, $t_2=0$, the effective Hamiltonian projected to $(A_1, B_4)$ sublattices is
\begin{equation}
H_{\rm 4LG}^{eff}(\vec k) = 
\begin{pmatrix}
\frac{3}{2}V_z & -\frac{v_0^4 k^4}{t_1^3} e^{-i4\theta_{\vec k}} \\
-\frac{v_0^4k^4}{t_1^3} e^{i4\theta_{\vec k}} & -\frac{3}{2}V_z
\end{pmatrix}.
\end{equation}

The conduction band Berry curvature of the above Hamiltonian is negative. The converged BCS Hamiltonian for the favored superconducting state is
\begin{equation}
        H^{\rm BCS}_c=
\begin{pmatrix}
\xi_{c,\vec k} & |\Delta_\vec k| e^{-i\phi_k}  \\
 |\Delta_\vec k| e^{i\phi_k} & -\xi_{c,-\vec k}
\end{pmatrix},
\end{equation}
a $p-ip$ chirality,  consistent with the analysis of in Ref.~\cite{may2025pairing}.

\section{Details of the Mean-Field Theory and Collective-Mode Calculations}

In this appendix, we provide the details behind our self-consistent mean-field calculations and how to used their result to set up a microscopic time-dependent mean-field calculation for collective modes. We assume the attractive interaction is much smaller than the spectral separation between the first conduction band to other bands and start from the following Hamiltonian,
\begin{gather}
H_c = \sum_{k}\xi_{c \vec k}
a^\dagger_{c\vec k} a_{c\vec k}+
\frac{1}{2} \sum\limits_{\vec k,\vec k'} \tilde{V}_{\vec k,\vec k'} a^\dagger_{c\vec k} a^\dagger_{c,-\vec k} a_{c,-\vec k'} a_{c\vec k'},\\
\tilde{V}_{\vec k,\vec k'} = -\frac{2g}{A}  \left( \cos(\phi_{\vec k}-\phi_{\vec k'}) \right)
\braket{\phi_{c, \vec k}|\phi_{c,\vec k'}}
    \braket{\phi_{c, -\vec k}|\phi_{c,-\vec k'}}.
\end{gather}
In the mean-field approximation,
\begin{equation}
    H_{c}^{\rm BCS}(\vec k) =
\begin{pmatrix}
\xi_{c\vec k} & \Delta_{\vec k} \\
\Delta_{\vec k}^* & -\xi_{c,-\vec k} \\
\end{pmatrix}
\end{equation}
and the zero-temperature self-consistent gap equation is given by 
\begin{equation}
\begin{split}
\Delta_{\vec k}
&= -\sum_{\vec k'} \tilde{V}_{\vec k,\vec k'} \frac{\Delta_{\vec k'}}{2\sqrt{(\xi_{c,\vec k'}^{e})^2 + |\Delta_{\vec k'}|^2}}
\end{split}
\end{equation}
with $\xi_{c\vec k}^e=(\xi_{c,\vec k}+\xi_{c,-\vec k})/2$. We seed $e^{i\phi_k}$ and the above equation will converge to $p+ip$ while  $e^{-i\phi_k}$, the gap solution will converge to  $p-ip$. Their degeneracy are lifted by sublattice winding.

The  eigenvectors and eigenenergies of the Bogoliubov quasiparticles take the following standard form,
\begin{equation}
\begin{split}
|U_{c,-,\vec k}\rangle &= [ \cos(\theta_{\vec k}/2), e^{i\phi_k}\sin(\theta_{\vec k}/2)]^T\\
|U_{c,+,\vec k}\rangle &=[\sin(\theta_{\vec k}/2), -e^{i\phi_k}\cos(\theta_{\vec k}/2) ]^T,\\
\end{split}
\end{equation}

\begin{equation}
E_{c,\pm,\vec k} = \xi^{o}_{c,\vec{k}} \pm\sqrt{(\xi^{e}_{c,\vec{k}})^2 + |\Delta_{\vec k}|^2}, 
\end{equation}

The BCS ground state can be described by a quasiparticle density-matrix made up by projectors to the lower Bogoliubov band $|U_{c,-,\vec k}\rangle$. 
\begin{equation}
    P_c=\sum_{\vec k}|U_{c,-,\vec k}\rangle\langle U_{c,-,\vec k}|
\end{equation}
One could loosely say that the lower Bogoliubov band is occupied while the upper
Bogoliubov band $|U_{c,+,\vec k}\rangle$ is empty. The subscript $c$ below $P_c$ emphasizes we are only focusing on the conduction band here, the full theory described in the main text contains all the bands. 

Now, we study the response of the ground state induced by the vector-potential,
\begin{equation}
    H'= e\Gamma^0_{\vec k}\cdot \vec A =e\begin{pmatrix}
      \vec v_{cc}(\vec k) & 0 \\
      0 & -\vec v_{cc}(-\vec k)
      \end{pmatrix}\cdot  \vec A
\end{equation}

This perturbation is going to generate transitions from the occupied $\ket{U_{c,-,\vec k}}$ to the unoccupied   $\ket{U_{c,+,\vec k}}$. The matrix element of this procdess is given by,
\begin{equation}
\vec{\gamma}_\vec{k}^0\equiv\bra{U_{c,+,\vec k}}\Gamma^0_{\vec k}\ket{U_{c,-,\vec k}}
    = \frac{\vec v_{cc}(\vec k)+\vec v_{cc}(-\vec k)}{2}
      \,\sin\theta_{\vec k}.
\end{equation}
In the presence of (spinless) time-reversal or inversion symmetry, $\vec v_{cc}(-\vec k)=-\vec v_{cc}(\vec k)$ and the above term vanish.  This perturbation changes the groundstate density matrix $P_c\rightarrow P_c+\delta P_c$. Using many body perturbation theory \cite{thouless2013quantum}, the matrix elements of the perturbed density matrix
$\braket{U_{c,+,\vec k}|\delta P_c|U_{c,-,\vec k}}\equiv X_{\vec k}$ can be obtained by solving the following equation:
\begin{equation} \label{eq:linearized_MF_supp_matt}
    \Delta E_\vec k\,X_{\vec k} + \delta \Sigma_{\vec k}
    = -e\vec{\gamma}_\vec{k}^0\cdot \vec A.
\end{equation}
Here $\Delta E_{\vec k}=E_{c,+,\vec k}-E_{c,-,\vec k}$ is the energy gap between the Bogoliubov bands. They 
 correspond to the energy required to create two non-interacting Bogoliubov quasiparticles with zero center-of-mass momentum on top of the BCS vacuum. In general, however, Bogoliubov quasiparticles do interact, and their interactions are decribed by the fluctuation of the pairing self-energy  $\delta \Sigma_{\vec k}$ in the time-dependent mean-field theory. To close this equation, we express $\delta \Sigma_{\vec k}$ in terms of the unknowns $X_\vec{k}$.  $\delta \Sigma_{\vec k}$ arises from fluctuations of the pairing potential $\delta \Delta_{\vec k}$, which are themselves generated
by fluctuations of the anomalous pairing amplitude $\delta F_{\vec k}$, and $\delta F_{\vec k}$ is related to $X_\vec k$.  They are described by the following equations,
\begin{align}
    \delta \Sigma_{\vec k}
    &=
    (a_{\vec k}^x + i\,b_{\vec k}^x)\delta \Delta_{\vec k}^R
    -(a_{\vec k}^y + i\,b_{\vec k}^y)\,\delta \Delta_{\vec k}^I, \\
    \delta \Delta_{\vec k}^R
    &=
    \sum_{\vec k'}
    \left(
        \tilde V^R_{\vec k\vec k'}\,\delta F_{\vec k'}^R
        - \tilde V^I_{\vec k\vec k'}\,\delta F_{\vec k'}^I
    \right), \\
    \delta \Delta_{\vec k}^I
    &=
    \sum_{\vec k'}
    \left(
        \tilde V^R_{\vec k\vec k'}\,\delta F_{\vec k'}^I
        + \tilde V^I_{\vec k\vec k'}\,\delta F_{\vec k'}^R
    \right), \\
    \delta F_{\vec k}^R
    &=X_{\vec k}^R a^x_ \vec k 
    + X_{\vec k}^I b^x_\vec k\\
    %=X_{\vec k'}^R\,\Re\!\left[\tau^x_{+-}(\vec k')\right]+ X_{\vec k'}^I\,\Im\!\left[\tau^x_{+-}(\vec k')\right], \\
    \delta F_{\vec k'}^I
    &=-X_{\vec k}^R a^y_{\vec k} 
    - X_{\vec k}^I b^y_{\vec k}
    %=-X_{\vec k'}^R\,\Re\!\left[\tau^y_{+-}(\vec k')\right] - X_{\vec k'}^I\,\Im\!\left[\tau^y_{+-}(\vec k')\right].
\end{align}
Combining these relations, we can express $\delta \Sigma_\vec k$ in terms of $X_\vec k$:
\begin{equation}\label{eq:Sigma}
    \begin{pmatrix}
    \delta \Sigma_\vec k^R\\  \delta \Sigma_\vec k^I
    \end{pmatrix}=
       \sum_{\vec k'}
       \begin{pmatrix}
    C_{\vec k,\vec k'}^{RR} & C_{\vec k,\vec k'}^{RI}
    \\ C_{\vec k,\vec k'}^{IR} & C_{\vec k,\vec k'}^{II}
    \end{pmatrix}
       \begin{pmatrix}
    X_{\vec k'}^R\\ X_{\vec k'}^I
    \end{pmatrix}
\end{equation}
where
\begin{align}
C^{RR}_{\vec k\vec k'}
&= \tilde V^R_{\vec k\vec k'}\big(a_{\vec k}^x a_{\vec k'}^x + a_{\vec k}^y a_{\vec k'}^y\big)
 + \tilde V^I_{\vec k\vec k'}\big(a_{\vec k}^x a_{\vec k'}^y - a_{\vec k}^y a_{\vec k'}^x\big), \\[4pt]
C^{RI}_{\vec k\vec k'}
&= \tilde V^R_{\vec k\vec k'}\big(a_{\vec k}^x b_{\vec k'}^x + a_{\vec k}^y b_{\vec k'}^y\big)
 + \tilde V^I_{\vec k\vec k'}\big(a_{\vec k}^x b_{\vec k'}^y - a_{\vec k}^y b_{\vec k'}^x\big), \\[4pt]
C^{IR}_{\vec k\vec k'}
&= \tilde V^R_{\vec k\vec k'}\big(b_{\vec k}^x a_{\vec k'}^x + b_{\vec k}^y a_{\vec k'}^y\big)
 + \tilde V^I_{\vec k\vec k'}\big(b_{\vec k}^x a_{\vec k'}^y - b_{\vec k}^y a_{\vec k'}^x\big), \\[4pt]
C^{II}_{\vec k\vec k'}
&= \tilde V^R_{\vec k\vec k'}\big(b_{\vec k}^x b_{\vec k'}^x + b_{\vec k}^y b_{\vec k'}^y\big)
 + \tilde V^I_{\vec k\vec k'}\big(b_{\vec k}^x b_{\vec k'}^y - b_{\vec k}^y b_{\vec k'}^x\big).
\end{align}
More compactly,
\begin{equation}
 C_{\vec k \vec k'}
=
\tilde V^{R}_{\vec k \vec k'}
\begin{pmatrix}
\mathbf a_{\vec k}\!\cdot\!\mathbf a_{\vec k'} &
\mathbf a_{\vec k}\!\cdot\!\mathbf b_{\vec k'} \\
\mathbf b_{\vec k}\!\cdot\!\mathbf a_{\vec k'} &
\mathbf b_{\vec k}\!\cdot\!\mathbf b_{\vec k'}
\end{pmatrix}
+
\tilde V^{I}_{\vec k \vec k'}
\begin{pmatrix}
\mathbf a_{\vec k}\!\times\!\mathbf a_{\vec k'} &
\mathbf a_{\vec k}\!\times\!\mathbf b_{\vec k'} \\
\mathbf b_{\vec k}\!\times\!\mathbf a_{\vec k'} &
\mathbf b_{\vec k}\!\times\!\mathbf b_{\vec k'}
\end{pmatrix}.
\end{equation}
Substituting Eq.~\eqref{eq:Sigma} into Eq.~\eqref{eq:linearized_MF_supp_matt}, we obtain the matrix equation discussed in the main text, Eq.~\eqref{eq:S_matrix}.

The above matrix equations are defined by 3 quantities: the Bogoliubov energy band gap $\Delta E_{\vec k}$, the real and imaginary parts of the interaction matrix element $\mathcal{V}^{R/I}$, and two real vector fields $\vec{a_k}=(a_{\vec k}^x,a_{\vec k}^y)$ and $\vec{b_k}=(b_{\vec k}^x,b_{\vec k}^y)$. The latter are defined through the matrix elements
\begin{align}
    \bra{U_{c,+,\vec k}}\tau^x\ket{U_{c,-,\vec k}}
    = a_{\vec k}^x + i\,b_{\vec k}^x, \\
    \bra{U_{c,+,\vec k}}\tau^y\ket{U_{c,-,\vec k}}
    = a_{\vec k}^y + i\,b_{\vec k}^y,
\end{align}
and admit a simple geometric interpretation. For a mean-field Hamiltonian, $H_c^{\rm{BCS}}(\vec{k})
= \xi_{c,\vec k}^o- \vec d_\vec k \cdot \vec \tau$, the local pseudospin direction is given by $\hat d_{\vec k}=\vec d_\vec k /|\vec d_\vec k |$. The two unit vectors
$\hat e_1(\vec k)$ and $\hat e_2(\vec k)$ span the tangent plane of the Bloch
sphere at $\hat d_{\vec k}$.
The vectors $\vec a_{\vec k}$ and $\vec b_{\vec k}$
are simply the projections of these tangent vectors onto the $xy$-plane of the global Nambu pseudospin basis $\tau=(\tau_x,\tau_y,\tau_z)$.  Explicitly, 
\begin{equation}
    \vec a_{\vec k} = P_{xy}\,\hat e_1(\vec k),
    \qquad
    \vec b_{\vec k} = P_{xy}\,\hat e_2(\vec k),
\end{equation}
where the projection operator is,
\begin{equation}
    P_{xy} =
    \begin{pmatrix}
        1 & 0 & 0 \\
        0 & 1 & 0 \\
        0 & 0 & 0
    \end{pmatrix}.
\end{equation}

A common approach to analyzing collective modes (e.g.~\cite{sauls2015anisotropy}) is to expand the order parameter in basis functions of an irreducible representation of the crystal point group and study the resulting Ginzburg–Landau dynamics in the long-wavelength limit. Our approach to computing the response of a superconductor follows Ref.~\cite{anderson1958random} and amounts to applying the linearized spin-wave calculation in magnetism to pseudospins in momentum space. This approach is particularly well suited to our continuum model which is not sensitive to point-group symmetry of the lattice and naturally incorporates the important bandstructure and form-factor details near the Brillouine zone corner. In this approach, we solve for the dynamics of pseudospin fluctuations generated by the mean-field pseudomagnetic field $\vec{B}_{\vec k}^{0}$ and its dynamical variation $\delta \vec{B}_{\vec k}$:
\begin{equation}
    \frac{d(\delta \vec{d}_{\vec k})}{dt}
    =
    \vec{B}_{\vec k}^{0} \times \delta \vec{d}_{\vec k}
    +
    \delta \vec{B}_{\vec k} \times \vec{d}_{\vec{k}}^{0}.
\end{equation}
In the conventional BCS ground state, the pseudospin texture $\vec{d}_{\vec{k}}^{0}$ forms a Bloch-type domain wall in momentum space \cite{anderson1958random}. 
By contrast, in our chiral topological p-wave superconductor the ground-state pseudospin texture $\vec{d}_{\vec{k}}^{0}$ forms a Skyrmion in momentum space. The collective modes studied here therefore correspond to excitations of this Skyrmion texture.
Unlike problems involving real spin, where the full tangent vectors $\hat e_1(\vec k)$ and $\hat e_2(\vec k)$ generally couple to each other. In superconductors, only the part of the tangent vectors that lie on the $xy$ plane, $\vec a_{\vec k}=P_{xy}\hat e_1(\vec k)$ and $\vec b_{\vec k}=P_{xy}\hat e_2(\vec k)$ enter the pseudospin dynamics. This is because only the  $\tau^x$ and $\tau^y$ component of the pseudospin leads to mixing between electrons and holes, i.e., pairing.

\begin{figure*}[t]
    \centering
    \includegraphics[width=1.0\linewidth]{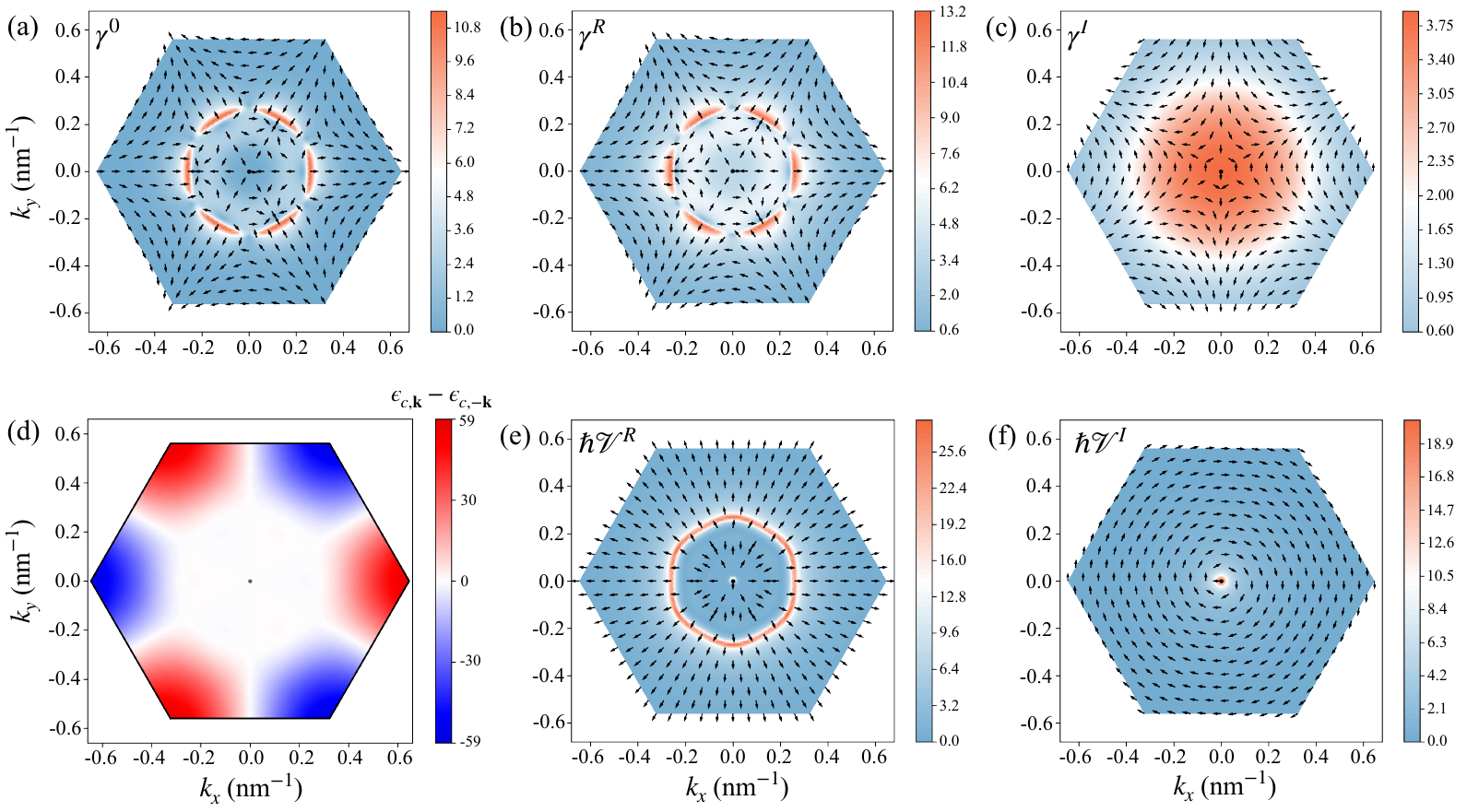}
    \caption{(a) Bare vertex $\gamma^0_{\bf{k}}$; (b-c) renormalized vertices $\gamma^R_{\bf{k}}$ and $\gamma^I_{\bf{k}}$; (e-f) quasiparticle velocities $\hbar \mathcal{V}^R$ and $\hbar \mathcal{V}^I$. In (a-c, e-f), arrows indicate the direction of the normalized vector fields, and color represents their magnitude. (d) Energy factor $(\epsilon_{c,\vec k}-\epsilon_{c,-\vec k})$ in he formula of for $M^{BB}$ in unit of meV. All panels correspond to the case studied in Figs.~\ref{fig:mf_solution}-\ref{fig:collective_modes}.} 
\label{figS_gamma}
\end{figure*}

The pseudospin interaction described by $C$ is quite illuminating. It  depends  on the relative orientation of the 
xy-projected tangent vectors $(\vec{a}_\vec k,\vec b_\vec k)$ at momentum $\vec k$ and  $(\vec{a}_\vec k',\vec b_\vec k')$ at momentum $\vec k'$. 
The real part of the pairing potential $ \tilde V^R_{\vec k\vec k'}$ multiplies a matrix composed of dot products between the these 
xy-projected tangent vectors. Thus, when the two xy-projected tangent vectors at  $\vec k$ and $\vec k'$ are nearly aligned, they experience a strong $\tilde V^R$. 
In contrast, the imaginary part of the pairing potential $\tilde V^I$ multiplies a matrix of cross products. This means when the xy-projected tangent vectors at  $\vec k$ and $\vec k'$ are nearly orthogonal, they experience a strong $\tilde V^I$.  Thus, depending on the strength of $\tilde V^R_{\vec k\vec k'}$ vs $\tilde V^I_{\vec k\vec k'}$, the tangent vectors  at $k$ and $k'$, preferentially align or become orthogonal. We note time-reversal symmetry requires $\tilde{V}^I=0$, 
so that orthogonal tangent-plane deformations do not couple in a time-reversal-invariant superconducting state. The cross-product structure multiplying  $\tilde{V}^I$ is  structurally analogous to a Dzyaloshinskii–Moriya interaction, but acting in momentum space on transverse pseudospin fluctuations rather than on real-space spins.

%This completes the derivation of the time-dependent mean-field equation for superconductors. 

Let us note some property of this matrix. Since $\tilde V_{\vec k\vec k'}$ is Hermitian, the real part is even and imaginary part is odd:
$\tilde V^R_{\vec k\vec k'}=\tilde V^R_{\vec k'\vec k}$ and
$\tilde V^I_{\vec k\vec k'}=-\tilde V^I_{\vec k'\vec k}$.
The dot-product terms are even under $\vec k\leftrightarrow\vec k'$, while the
cross-product terms are odd, thus, $ C_{\vec k,\vec k'} = C_{\vec k',\vec k}$. As a result, the full stability matrix $\mathcal{S}$ in Eq.~\eqref{eq:S_matrix} is real and symmetric,
    $\mathcal{S}=\mathcal{S}^T$,
and all of its eigenvalues are real. The eigenvalues of $\mathcal{S}$ control the quadratic variation of the mean-field grand potential under deformations of the superconducting pseudospin texture:
$\delta^2\Omega
    = \tfrac{1}{2}
    \sum_{\vec k,\vec k'}
    \begin{pmatrix}
        X_{\vec k}^R & X_{\vec k}^I
    \end{pmatrix}
    \mathcal{S}_{\vec k,\vec k'}
    \begin{pmatrix}
        X_{\vec k'}^R \\
        X_{\vec k'}^I
    \end{pmatrix},
$
which is the analog of the Thouless stability matrix \cite{thouless2013quantum} in Hartree–Fock theory. For a locally stable mean-field solution, $\delta^2\Omega\geq0$, eigenvalues of $\mathcal{S}$ must be non-negative.

In order to obtain the dressed vertex, we substitute
$X_{\vec k} = -\,e\,\vec\gamma_{\vec k}\cdot\vec A/\Delta E_{\vec k}
$  and arrive at the following vertex equation:
\begin{equation}\label{eq:supp_matt_vertex}
    \sum_{\mathbf{k}'}
\begin{pmatrix}
 \delta_{\mathbf{k}\mathbf{k}'} + C^{RR}_{\mathbf{k},\mathbf{k}'}/\Delta E_{\vec k'} &
C^{RI}_{\mathbf{k},\mathbf{k}'}/\Delta E_{\vec k'}\\[4pt]
C^{IR}_{\mathbf{k},\mathbf{k}'}/\Delta E_{\vec k'} &
\delta_{\mathbf{k}\mathbf{k}'} + C^{II}_{\mathbf{k},\mathbf{k}'}/\Delta E_{\vec k'}
\end{pmatrix}
\begin{pmatrix}
\gamma^R_{\mathbf{k}'}\\[2pt]
\gamma^I_{\mathbf{k}'}
\end{pmatrix}
=
\begin{pmatrix}
\vec{\gamma}^{\,0}_{\mathbf{k}}\\
0
\end{pmatrix}.
\end{equation}

The bare photon vertex $\vec{\gamma}^{\,0}_{\mathbf{k}}$, together with the real ($\gamma^R_{\mathbf{k}}$) and imaginary parts ($\gamma^I_{\mathbf{k}}$) of the dressed vertex field, are shown in Fig.~\ref{figS_gamma}. We observe that the real part of the dressed vertex is only weakly renormalized and is mainly concentrated near the Fermi surface. By contrast, the imaginary part, which arises entirely from vertex corrections, is pronounced in the entire Fermi sea.  In Fig.~\ref{figS_gamma}, we also plot the quasiparticle velocity matrix elements,
\begin{equation}
\frac{1}{\hbar}\langle U_{c,- , \vec k} | \nabla H^{\rm BCS}_c(\vec k) | U_{c,+ , \vec k} \rangle
\equiv \mathcal{V}^R_{\vec k} + i \mathcal{V}^I_{\vec k},
\end{equation}
which define two vector fields corresponding to their real and imaginary parts. $\mathcal{V}^R$ is peaked at the Fermi surface, similar to $\gamma^R$, although its vector-field structure is completely different. The imaginary part $\mathcal{V}^I$ has a vortex-like structure and is peaked inside the Fermi sea, where the pseudospin changes rapidly, and is likewise distinct from $\gamma^I$.
Unlike in the normal state, the quasiparticle velocity fields in the superconducting phase are distinct from the photon vertex fields. Both the photon vertex and the quasiparticle velocity fields are required to determine the intrinsic orbital magnetization generated by Cooper pairs in the superconductor:
\begin{equation} 
\begin{split}\label{eq:supp_matt_M_BB}
M_z^{\rm BB}(\vec k) = 
\frac{e\hbar}{8} 
\left(\frac{\epsilon_{c,\vec k}-\epsilon_{c,-\vec k}}{2}\right)
\frac{\mathcal{V}^I_\vec{k} \times \gamma^R_\vec{k}+
\mathcal{V}^R_\vec{k} \times 
\gamma^I_\vec{k}  }{(\xi_{c,\vec k}^{e})^2 + |\Delta_{\vec k}|^2}.
\end{split}
\end{equation}

\section{Additional collective mode}
\begin{figure}[t]
    \centering
    \includegraphics[width=1.0\linewidth]{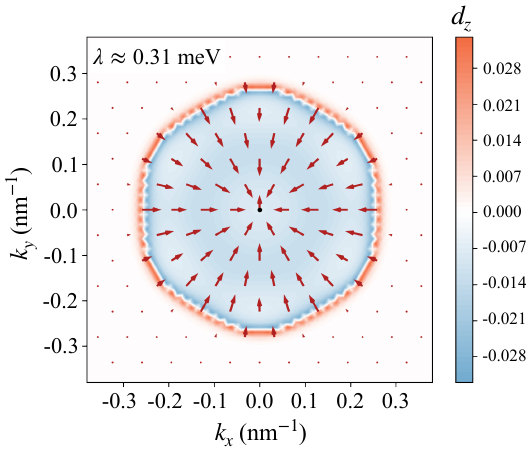}
    \caption{Pseudospin texture of the collective mode at $\lambda \approx 0.31$~meV indicated in Fig.~\ref{fig:collective_modes}(a). Arrows denote the in-plane pseudospin components $(d_x, d_y)$. The vector field is shown without normalization to emphasize the collective nature of the excitation. Calculation is done for valley $K$ rhombohedral tetralayer graphene at $V_z=41$~meV and carrier density $n_e=0.5\times10^{12}$~cm$^{-2}$.}
\label{figS:higgs}
\end{figure}

In Fig.~\ref{fig:collective_modes}(a) of the main text, in addition to the Goldstone mode and the twofold-degenerate generalized clapping modes, we identify an additional collective mode lying below the quasiparticle continuum at $\lambda \approx 0.31$~meV. The pseudospin texture of this mode is shown in Fig.~\ref{figS:higgs}. This mode carries the same $p\!-\!ip$ winding as the ground state and therefore cannot be identified as a clapping mode.
We do not analyze this mode in detail because it is charged, $\sum_{\vec k} d^z_{\vec k} \neq 0$. While this mode contribute to the dressing of the photon vertex, it is expected to be strongly hybridize with the plasmon in optical probes \cite{sun2020collective}, making it difficult to isolate experimentally.

\newpage
\section{Data at other density--displacement-field points}
In the main text, we focused on a representative density–displacement-field ($n$–$D$) point that has a single Fermi surface with trigonal warping. We analyzed the superconducting band structure (Fig.~2), the different contributions to the orbital magnetization (Fig.~3), and the collective-mode spectrum and associated pseudospin textures (Fig.~4). Here, we present the corresponding results at other $n$–$D$ points.

%\begin{figure*}[t]
%    \centering   \includegraphics[width=1.0\linewidth]{figS1.pdf}   \caption{$k$-space distribution of (a-b) the energy factor $(E_{c,+,\bf{k}} + E_{c,-,\bf{k}})/2$, the colorbar of (b) is truncated to see the detailed structure. (c) Berry-curvature-like factor $\epsilon_{\alpha \beta} \pi \text{Im} [\langle \Gamma^\beta_{\bf k}\rangle_{+-} \langle \partial_{\alpha}H_c^{BCS}({\bf k}) \rangle_{-+}]/[(E_{c,+,{\bf k}} - E_{c,-,{\bf k}})^2 \phi_0]$ which is in unit of Tesla$^{-1}$. Here $\langle \cdot \rangle_{+-}$ denotes $\langle U_{c,+,\bf{k}}|\cdot|U_{c,-,\bf{k}}\rangle$.    (d) $M_z^{BB}(k)$ in unit of $\mu_B$ per electron.    From (a-c), the small value of $M_z^{BB}$ is a direct consequence of the vanishingly small energy factor near and inside the Fermi surface. (a-d) for valley $K$ at $V_z=41$~meV and carrier density $n_e=0.5\times10^{12}$~cm$^{-2}$, $g=160$~meV$\cdot$nm$^2$.    (e-h) for valley $K$ at $V_z=45$~meV and carrier density $n_e=0.2\times10^{12}$~cm$^{-2}$, $g=160$~meV$\cdot$nm$^2$.} \label{figS1:MzBB_contributions}
%\end{figure*}

\begin{figure*}[t]
    \centering
    \includegraphics[width=1.0\linewidth]{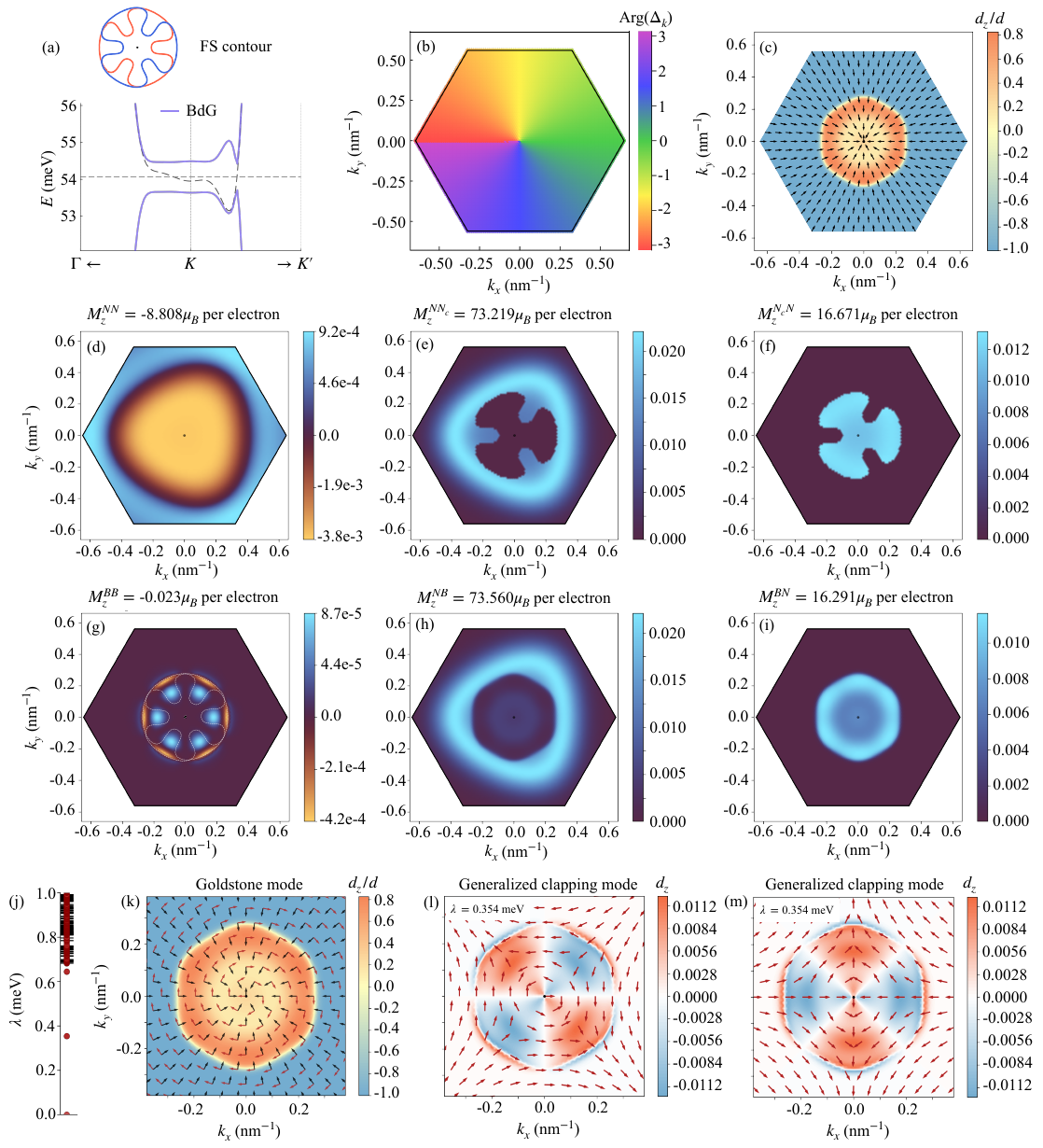}
    \caption{
    Superconducting mean-field, the different contributions to the orbital magnetization, and the collective-mode spectrum and associated pseudospin textures at $V_z=43$~meV and carrier density $n_e=0.45\times10^{12}$~cm$^{-2}$, corresponding to the lowest-density point shown in Fig.~\ref{fig:schematic}b).} 
\label{figS2}
\end{figure*}

\begin{figure*}[t]
    \centering
    \includegraphics[width=1.0\linewidth]{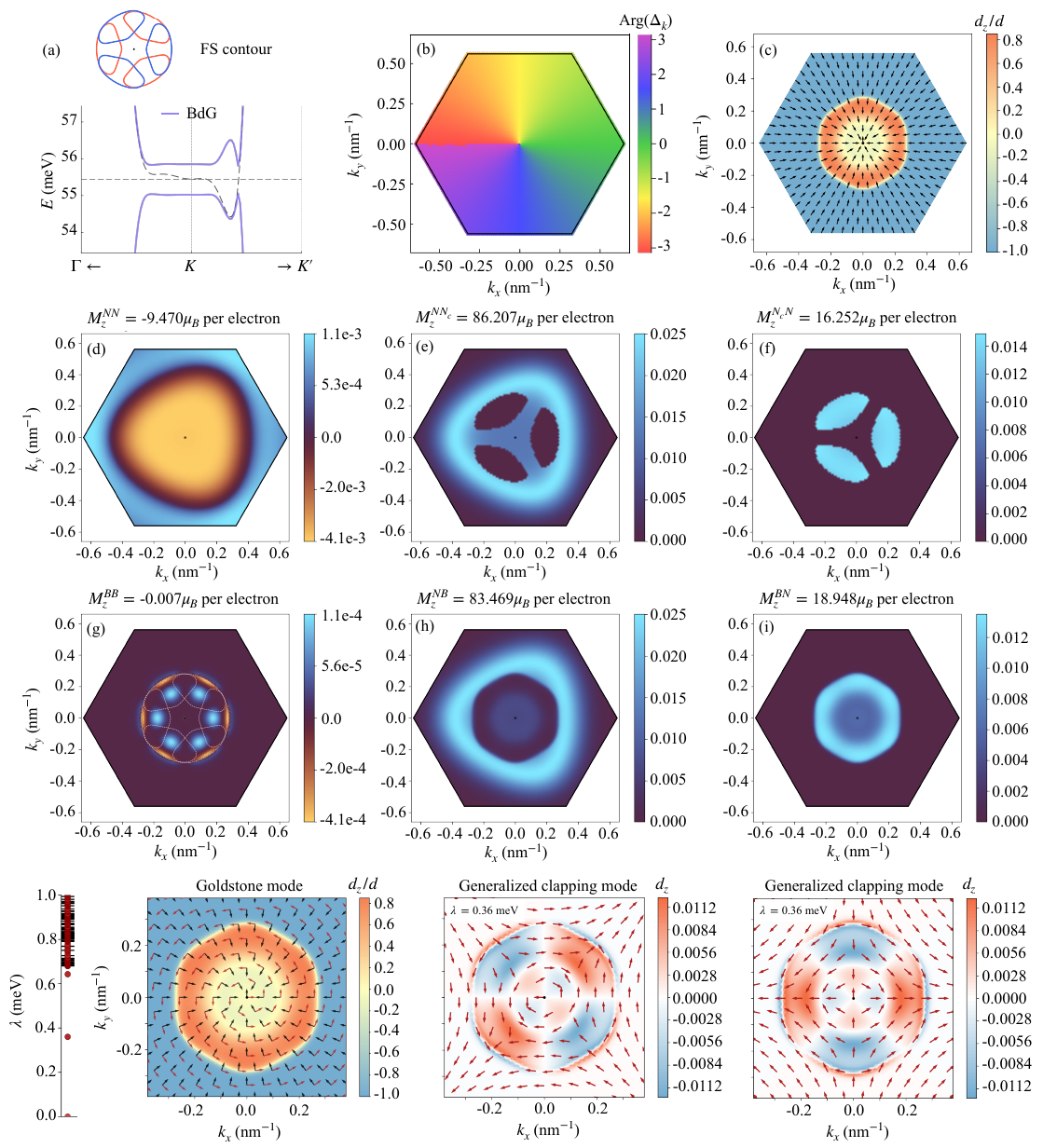}
    \caption{Superconducting mean-field, the different contributions to the orbital magnetization, and the collective-mode spectrum and associated pseudospin textures at $V_z=44$~meV and carrier density $n_e=0.4\times10^{12}$~cm$^{-2}$.} 
\label{figS3}
\end{figure*}

\begin{figure*}[t]
    \centering
    \includegraphics[width=1.0\linewidth]{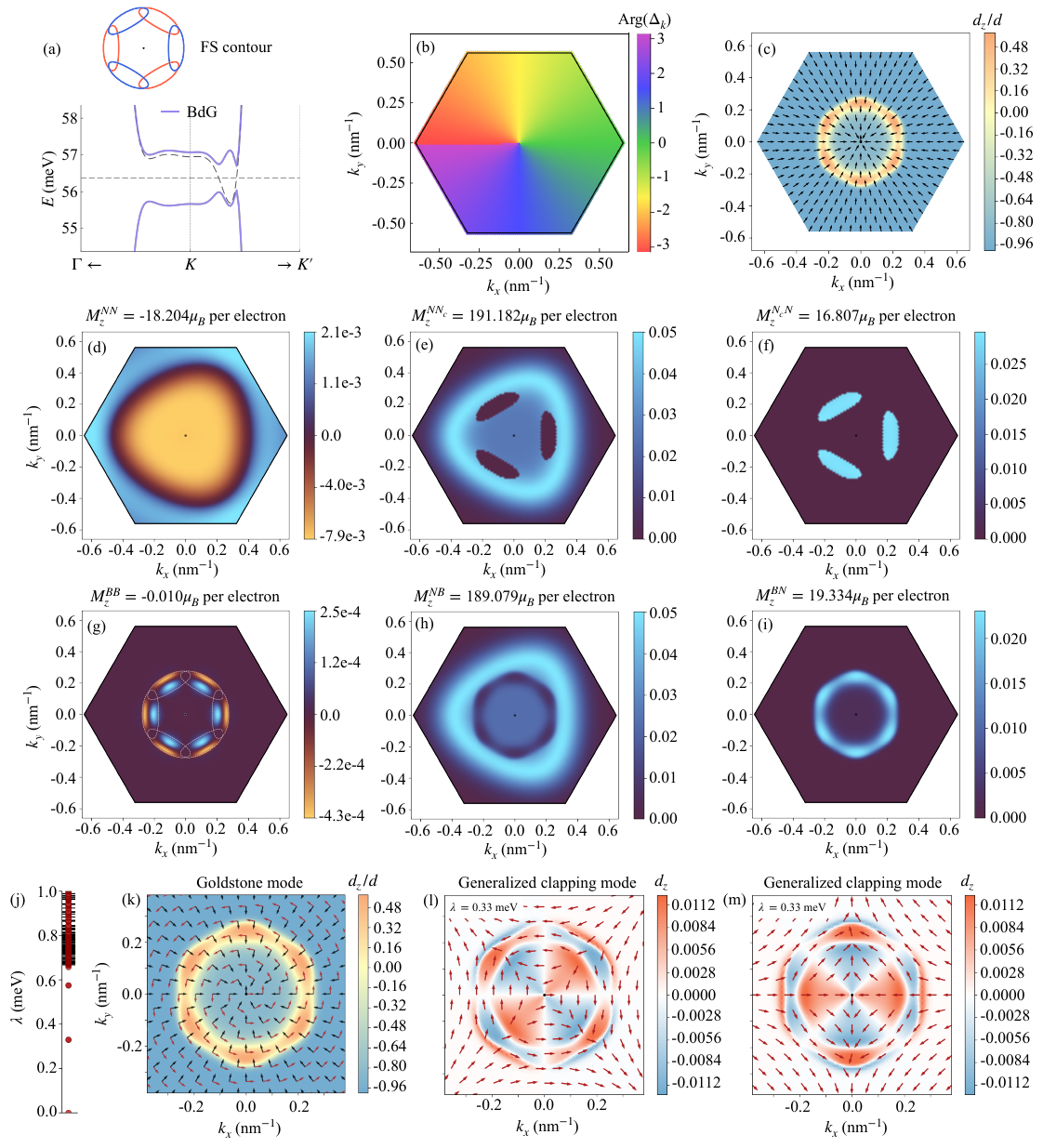}
    \caption{Superconducting mean-field, the different contributions to the orbital magnetization, and the collective-mode spectrum and associated pseudospin textures at at $V_z=45$~meV and carrier density $n_e=0.2\times10^{12}$~cm$^{-2}$.} 
\label{figS4}
\end{figure*}

\begin{figure*}[t]
    \centering
    \includegraphics[width=1.0\linewidth]{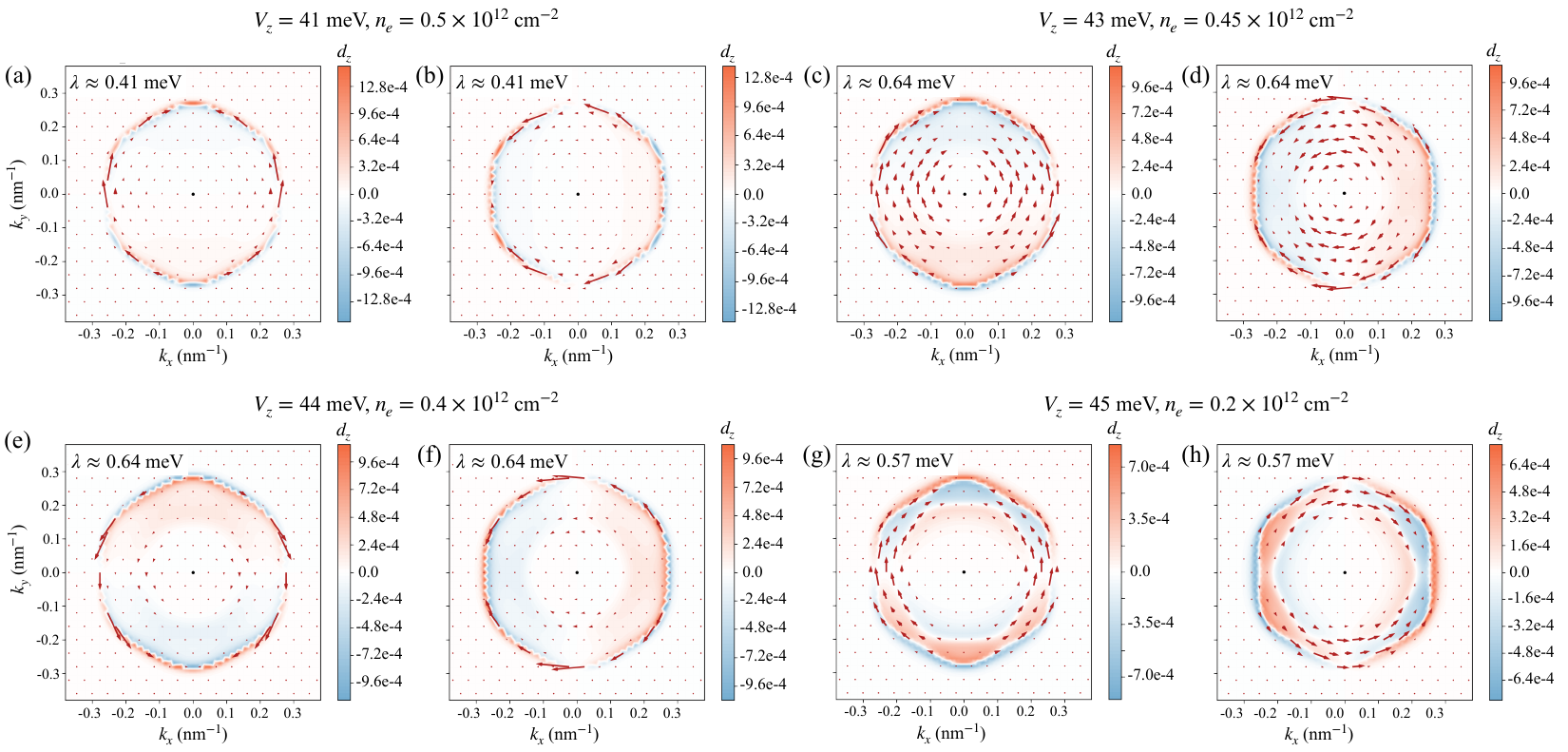}
    \caption{Additional modes with energies lower than the continuum. (a-b) For $V_z=41$~meV and $n_e=0.5\times10^{12}$~cm$^{-2}$, corresponding to the case in Figs.~\ref{fig:collective_modes} and \ref{figS:higgs}. These two degenerate charge-neutral collective modes have slightly higher energy than the charged mode in Fig.~\ref{figS:higgs}. (c-d) $V_z=43$~meV and $n_e=0.45\times10^{12}$~cm$^{-2}$, corresponding to the case in Fig.~\ref{figS2}. (e-f) $V_z=44$~meV and $n_e=0.4\times10^{12}$~cm$^{-2}$, corresponding to the case in Fig.~\ref{figS3}. (g-h) $V_z=45$~meV and $n_e=0.2\times10^{12}$~cm$^{-2}$, corresponding to the case in Fig.~\ref{figS4}. The modes shown in panels (c), (d), (g), and (h) correspond to collective excitations characterized by two nodes in the superconducting order parameter.} 
\label{figS5}
\end{figure*}

\end{document}